\documentclass[journal]{IEEEtran}
\usepackage{graphicx} 
\usepackage{amsmath,amssymb,amsfonts,amsthm}
\usepackage{multicol,lipsum}
\usepackage{mathtools}
\usepackage{cite}
\usepackage{algorithmic}
\usepackage{algorithm}
\usepackage{enumerate}
\usepackage{comment}

\DeclareMathOperator{\Tr}{Tr}
\DeclareMathOperator{\Vect}{vec}

\newtheorem{lemma}{\quad \textit{Lemma}}
\newtheorem{remark}{\quad \textit{Remark}}

\usepackage{subcaption}
\usepackage{graphicx}
\usepackage{textcomp}
\usepackage{optidef}
\usepackage{xcolor}

\title{Fluid Antenna-enabled Near-Field Integrated Sensing, Computing and Semantic Communication for Emerging Applications}

\begin{document}
\author{Yinchao Yang, Jingxuan Zhou,
Zhaohui Yang, Mohammad Shikh-Bahaei,
\IEEEmembership{Senior Member, IEEE}
\thanks{Yinchao Yang, Jingxuan Zhou, and Mohammad Shikh-Bahaei are with the Department of Engineering, King's College London, London, UK. (emails: yinchao.yang@kcl.ac.uk; jingxuan.zhou@kcl.ac.uk; m.sbahaei@kcl.ac.uk)}
\thanks{Zhaohui Yang is with the College of Information Science and Electronic Engineering, Zhejiang University, Hangzhou, Zhejiang 310027, China, and Zhejiang Provincial Key Lab of Information Processing, Communication and Networking (IPCAN), Hangzhou, Zhejiang, 310007, China. (email: yang\_zhaohui@zju.edu.cn) }}

\maketitle
\IEEEpeerreviewmaketitle
\begin{abstract}

The integration of sensing and communication (ISAC) has been recognised as a key enabler for next-generation technologies. Meanwhile, with the adoption of high-frequency bands and large-scale antenna arrays, the extension of the Rayleigh distance necessitates the consideration of near-field (NF) models, where signal waves are spherical. While NF-ISAC enhances both sensing and communication, it introduces significant challenges, including increased data volume and potential privacy concerns. To address these issues, this paper introduces a novel framework: near-field integrated sensing, computing, and semantic communication (NF-ISCSC), which incorporates semantic communication to transmit only contextual information, thereby reducing data overhead and improving system efficiency. However, the sensitivity of semantic communication to channel conditions underscores the need for adaptive solutions. To this end, fluid antennas (FAs) are proposed to assist the NF-ISCSC system, offering dynamic adaptability to channel variations. The proposed FA-enabled NF-ISCSC framework takes into account multiple communication users, and extended targets which are composed of a series of scatterers. A joint optimisation problem is formulated to maximise data rate while accounting for sensing performance, computational constraints, and the power budget. By using an alternating optimisation (AO) approach, the original optimisation problem is decomposed into three sub-problems to iteratively optimise ISAC beamforming, FA positioning, and semantic extraction ratio. The beamforming optimisation is solved using the successive convex approximation method. The FA positioning problem is solved via a computationally efficient projected Broyden–Fletcher–Goldfarb–Shann (projected BFGS) algorithm. The semantic extraction ratio optimisation employs a bisection search method. Simulation results validate the proposed framework, highlighting the significant benefits of achieving higher data rates and enhanced privacy.

\end{abstract}


\section{Introduction}

With the rapid growth of emerging applications such as virtual/augmented reality (VR/AR) and digital twins (DTs), the integration of wireless sensing and communication is ready to play a pivotal role in enabling next-generation services \cite{olfat2008optimum, shikh2007joint, bobarshad2010low, nehra2010cross, bobarshad2009m, nehra2010spectral, jia2020channel, kobravi2007cross, shadmand2010multi, towhidlou2017improved, shojaeifard2011joint, yang2025toward, yang2025integrated}. Recognizing its transformative potential, the International Telecommunication Union (ITU) has incorporated integrated sensing and communication (ISAC) as one of the six core usage scenarios in its global vision for the sixth-generation (6G) mobile communication systems.

The basic principle of ISAC is to deploy wireless sensing and communication capability through the shared use of wireless resources, including power, frequency bands, beams, and hardware infrastructure. Numerous studies, to list a few, \cite{dong2022sensing, yuan2020bayesian, su2020secure, lu2024random}, have demonstrated that ISAC systems can outperform standalone sensing and communication systems in terms of resource efficiency and can unlock synergistic advantages, where sensing enhances communication and communication supports sensing. On the one hand, sensing-assisted communication uses sensing data to provide contextual awareness, which helps optimize communication processes like beamforming, channel estimation, and resource allocation by inferring channel quality and proactively modifying the system's settings \cite{liu2022integrated}. On the other hand, in a communication-assisted sensing framework, communication devices support sensing by transmitting sensory data and capturing reflected signals, eliminating the need for dedicated sensors. This simplifies network architecture and reduces hardware costs compared to traditional sensor-based networks \cite{zhang2021overview}.

Despite significant advancements in ISAC, meeting the exceptionally high data rate requirements and supporting massive connectivity for emerging applications remains a critical challenge. These demands necessitate the deployment of massive antenna arrays or the use of extremely high-frequency bands \cite{liu2023near}. Specifically, when the distance between the transmitter and receiver is less than $\frac{2D^2}{\lambda}$ (i.e. the Rayleigh distance), where $D$ represents the array aperture and $\lambda$ is the wavelength, a near-field (NF) model becomes essential to accurately capture the spherical wave propagation characteristics \cite{babu2024symbol}. For instance, with a transmitter aperture of $D = 0.5 \;\text{m}$ operating at a frequency of $50\; \text{GHz}$, the NF region extends to $83\;\text{m}$.

Several key advantages emerge when considering the NF model. First, unlike far-field (FF) beamforming, which directs beam energy along a fixed path, NF beam focusing utilises spherical wavefronts to concentrate energy at a specific location \cite{cui2022near, zhang2022beam}. This focusing capability enhances the received signal power for intended users. In terms of sensing capabilities, the NF model demonstrates superior performance in target localisation \cite{wang2024cramer}. The spherical wavefront properties enable simultaneous estimation of both angular and range information with high precision, particularly when employing advanced techniques such as multiple signal classification (MUSIC) with fine-grid processing. While Doppler-based methods in FF scenarios can estimate range through velocity measurements, those methods often yield less accurate results, especially in static or low-mobility scenarios where frequency shifts are minimal. Furthermore, the NF beam focusing effect provides enhanced control over the sensing signal distribution, effectively reducing interference in echo signals and improving overall sensing accuracy \cite{zhang2024physical}. This characteristic is particularly valuable in complex environments where multiple line-of-sight (LoS) targets or reflectors may be present. Several studies have explored the NF-ISAC system. For instance, Wang \textit{et al.} \cite{wang2023near} analysed NF-ISAC systems with multiple communication users and a single target, investigating both fully digital antenna arrays and hybrid digital-analogue configurations. Their findings highlight the performance enhancements enabled by exploiting the additional distance dimension in NF-ISAC compared to FF-ISAC. Zhao \textit{et al.} \cite{zhao2024modeling} examined the trade-off between communication and sensing functionalities in both downlink and uplink scenarios. Qu \textit{et al.} \cite{qu2024near} compared NF beam focusing with FF beam steering, demonstrating through simulations that NF beam focusing effectively mitigates co-angle interference while enhancing both communication and sensing performance.

While NF-ISAC offers promising advantages, it faces substantial challenges, particularly in managing the enormous volume of data transmitted and received in real-time scenarios such as autonomous driving. The excessive data volume can lead to increased latency and computational overhead, hindering system performance \cite{gu2023semantic}. Another critical challenge involves potential data privacy risks. As sensing and communication signals are transmitted simultaneously, sensing targets (which are unintended communication users) may inadvertently receive confidential communication messages, raising significant privacy concerns \cite{zhaohui2024secure}. The integration of semantic communication into the NF-ISAC framework presents a promising solution to these challenges. Semantic communication focuses on transmitting the meaning and relevance of information rather than raw data, enabling intelligent prioritisation and compression based on contextual importance \cite{luo2022semantic}. For example, in autonomous driving, instead of transmitting comprehensive sensing results, the system could extract and transmit semantically relevant information, such as the trajectories of critical objects or potential hazards, to other roadside units. This semantic-aware approach significantly reduces the data burden and enhances system efficiency by ensuring that limited resources are allocated to transmitting the most meaningful and actionable information. To enable semantic communication, both transmitters and receivers must establish shared knowledge bases (KBs) containing essential information accessible to both parties \cite{xu2023edge, gunduz2022beyond}. 
Crucially, the security and integrity of these KBs are central to maintaining the privacy and security of semantic communication. Regulations such as the General Data Protection Regulation (GDPR), which governs the handling of personal data within the EU and EEA, provide a legal and procedural framework for protecting sensitive data \cite{do2025security, agarwal2022gdpr}. By adhering to GDPR principles or similar regulatory frameworks, KBs can be safeguarded against unauthorised access and cyber-attacks \cite{guo2024survey, won2024resource}. As a result, semantic communication inherently enhances data privacy: since only high-level semantic representations are transmitted, and successful decoding requires access to valid KBs, unintended users are effectively excluded from recovering the original message. Moreover, semantic communication facilitates efficient multi-modal data transmission, meeting the diverse requirements of different communication users \cite{zhang2024unified}. Despite these advantages of semantic communication, existing research on semantic communication, such as \cite{lu2022rethinking, xie2022task, xie2021deep}, has largely overlooked its integration into the NF-ISAC framework. This research gap represents a significant opportunity for advancing NF-ISAC systems by leveraging the benefits of semantic communication to address data overload, privacy concerns, and multi-modal data requirements.

Nonetheless, integrating semantic communication into the NF-ISAC framework introduces specific challenges, particularly due to the interference caused by the sensing signals. Unlike classical communication, which focuses on bit-level accuracy, semantic communication aims to preserve the intended meaning or context of the transmitted information. This makes it inherently more sensitive to channel conditions and interferences. Under poor channel conditions, not only can the data become corrupted, but the semantic meaning of a message may also be distorted—a phenomenon referred to as semantic noise \cite{luo2022semantic}. Semantic noise occurs when the receiver misinterprets the message, deviating from the sender’s original intent, which can lead to misunderstandings or incorrect decisions. For example, suppose a sender intends to transmit the message ``Alice’s new bike was broken''. In a traditional communication system, a noisy channel might corrupt it to ``Alace’s new baki was brakon'', maintaining the structure but affecting intelligibility. In contrast, a semantic communication system might deliver “Alisa’s bike is broken,” preserving a plausible meaning but deviating from the exact intended context. Therefore, developing strategies to mitigate semantic noise is essential for maintaining reliability in semantic communication systems.

One promising technique to stabilise the channel condition is to use the fluid antennas (FAs), which are flexible in terms of their positions to reconfigure the radiation characteristics \cite{ghadi2024physical, zhou2024near,zhu2023movable}. Compared to fixed-position antennas (FPAs), the FAs have a greater degree of freedom (DoF) to explore the channel variations to adapt to the changing channel conditions. Various studies have highlighted the significant advantages of FAs in wireless communication systems. For instance, Zhu \textit{et al.} \cite{zhu2023modeling} compared the communication performance of FPAs and FAs by analysing the maximum channel gain achieved in deterministic and stochastic channel environments. Their findings show that FAs outperform FPAs, especially in scenarios with an increasing number of channel paths, due to their ability to exploit more pronounced small-scale fading effects in the spatial domain. Similarly, Wang \textit{et al.} \cite{wang2024fluid} explored the role of FAs in ISAC systems within a multi-user multiple-input multiple-output (MU-MIMO) downlink framework. In this study, a base station (BS) equipped with FAs demonstrated enhanced beamforming precision, reduced inter-user interference, and improved overall system capacity. These advancements illustrate the potential of FAs to optimise both communication and sensing performance in complex multi-user scenarios. Despite the substantial progress made in demonstrating the benefits of FAs for improving communication performance in conventional systems, their applications in semantic communication and NF-ISAC systems have received minimal attention.

In \cite{yang2024secure}, the framework of integrated sensing, computing, and semantic communication (ISCSC) was proposed. However, the impact of NF effects and the potential advantages of incorporating FAs into the system were not explored. This limitation highlights a significant research gap, motivating the development of \textbf{fluid antenna-enabled NF-ISCSC (FA-enabled NF-ISCSC)} systems. Compared to the ISAC design, the proposed framework achieves higher data rates without sacrificing sensing performance, or alternatively, enhances sensing accuracy while maintaining the same data rate. These advantages make it especially suitable for applications such as vehicular networks and digital twins. In vehicular networks, particularly within the intelligent transportation system (ITS), sensing enables roadside units (RSUs) and vehicles to acquire essential environmental information, such as traffic signs, road conditions, and dynamic obstacles \cite{yang2022semantic}. Semantic communication ensures that only task-relevant and context-specific information, such as the sudden appearance of a pedestrian, is exchanged among RSUs and between RSUs and vehicles. For example, rather than transmitting raw sensor data, an RSU can convey a high-level semantic message like ``vehicle merging from left at 25 m/s'', thereby reducing the communication load and improving the accuracy and timeliness of information delivery. The use of FAs further enhances communication reliability and adaptability by reducing semantic errors. Computing plays a critical role by processing sensed data for real-time scene understanding, including object detection, tracking, and trajectory prediction \cite{wen2024survey}. Additionally, computing enables semantic communication by extracting and prioritising key features from sensory data for efficient and meaningful transmission. In the context of digital twins, sensing data collected by end devices enables real-time perception of the physical entity’s state. In the uplink scenario, semantic communication allows the end device to interpret and filter out irrelevant information prior to the transmission, thereby reducing bandwidth consumption and minimising latency when uploading to the digital twin server \cite{jagatheesaperumal2023semantic}. In the downlink, the server can generate and transmit semantic information based on environmental context, thus alleviating downlink resource congestion. For example, in the context of a digital twin for eHealth, instead of transmitting the full message ``The patient is a 68-year-old male with a medical history of Type 2 diabetes mellitus, hypertension, and coronary artery disease. He presented to the emergency department with chest pain that radiated to his left arm'', the end device may send a high-level semantic interpretation such as ``68-year-old male with chronic cardiovascular and metabolic diseases presented with radiating chest pain to the ED''. In this semantically extracted message, the word ``chronic'' captures the notion of a long-standing medical history, Type 2 diabetes mellitus is classified as a metabolic disease, while hypertension and coronary artery disease are considered cardiovascular diseases. With the KBs store the patient's electronic health record (EHR) and medical abbreviations like ``ED'' for emergency department, the extracted message serves as a medically meaningful and compressed representation of the patient’s historical conditions, reducing both data volume and transmission delay. The integration of fluid antennas enhances semantic communication by mitigating semantic errors and ensuring accurate message interpretation. Additionally, computing plays a vital role in processing sensed data, simulating physical behaviours, and extracting high-level semantic features necessary for efficient and meaningful communication.

Building on the motivation, the key contributions of this paper are:
\begin{enumerate}
    \item \textbf{NF-ISCSC Modelling with FAs}:
    This paper proposes an FA-enabled ISCSC framework that incorporates the NF effects, a crucial consideration for next-generation wireless systems operating at ultra-high frequency bands. Compared to ISAC, the incorporation of semantic communication not only enhances communication performance but also strengthens data privacy. Furthermore, the deployment of FAs facilitates channel condition control, thereby supporting and improving semantic communication.

    \item \textbf{Accurate Target Modelling}:  
    This paper employs an extended target model in which each target is represented by multiple scatterers, providing a detailed characterisation of real-world physical properties. Such precise modelling is critical for applications like DTs, where accurate and high-fidelity object representations are indispensable for effective simulation and decision-making.

    \item \textbf{Computational Resource and Power Optimisation}: The proposed framework formulates a joint optimisation problem that integrates the allocation of computational resources and power consumption for sensing data processing with the power required for semantic information extraction. Additionally, it incorporates the optimisation of ISAC beamforming, FA positions, and semantic extraction. The objective is to maximise system throughput and enhance data privacy while maintaining optimal sensing accuracy. This is achieved under the constraints of limited power budgets and computational resources. 

    \item \textbf{Reduced-Complexity Algorithm}: To address the non-convex nature of the FA-enabled NF-ISCSC optimisation problem, the problem is decomposed into three sub-problems using an alternating optimisation (AO) approach: ISAC beamforming optimisation, FA positioning optimisation, and semantic extraction ratio optimisation. The beamforming optimisation is solved using successive convex approximation, which effectively simplifies the non-convex problem into a series of convex approximations. For FA positioning optimisation, a computationally efficient projected Broyden–Fletcher–Goldfarb–Shanno (projected BFGS) algorithm is proposed, coupled with a backtracking line search (BLS) to determine the optimal step size. Finally, semantic extraction optimisation is addressed through search methods such as the bisection method. 
    
\end{enumerate}

The remainder of this paper is organised as follows: Section II describes the system model for the FA-enabled NF-ISCSC system. Section III formulates the performance indicators for the system. Section IV discussed the problem formulation and the solutions. Simulation results are provided in Section V, and conclusions are drawn in Section VI.

\subsection*{List of Notations:}
Capital boldface letters (e.g., $\mathbf{A}$) denote matrices, while lowercase boldface letters (e.g., $\mathbf{a}$) denote vectors. Scalars are represented using regular lowercase or uppercase fonts. The sets $\mathbb{C}$, $\mathbb{C}^{n \times 1}$, and $\mathbb{C}^{m \times n}$ represent a complex number, a complex vector of length $n$, and a complex $m \times n$ matrix, respectively. The identity matrix is denoted by $\mathbf{I}$, and the zero matrix by $\mathbf{0}$. The notations $[\cdot]^H$, $\Tr(\cdot)$, $\Vect( \cdot)$, and $\text{rank}(\cdot)$ represent the Hermitian transpose, the trace, the vectorisation, and the rank of a matrix, respectively. The $l_2$ norm is denoted by $||\cdot||$. The symbol $\succeq$ indicates positive semi-definiteness, and $\otimes$ denotes the Kronecker product. Finally, $\mathcal{CN}(0, \sigma^2)$ represents a standard complex Gaussian distribution with zero mean and variance $\sigma^2$.

\section{System Model}

\begin{figure}[!t]
    \centering
    \includegraphics[width=\linewidth]{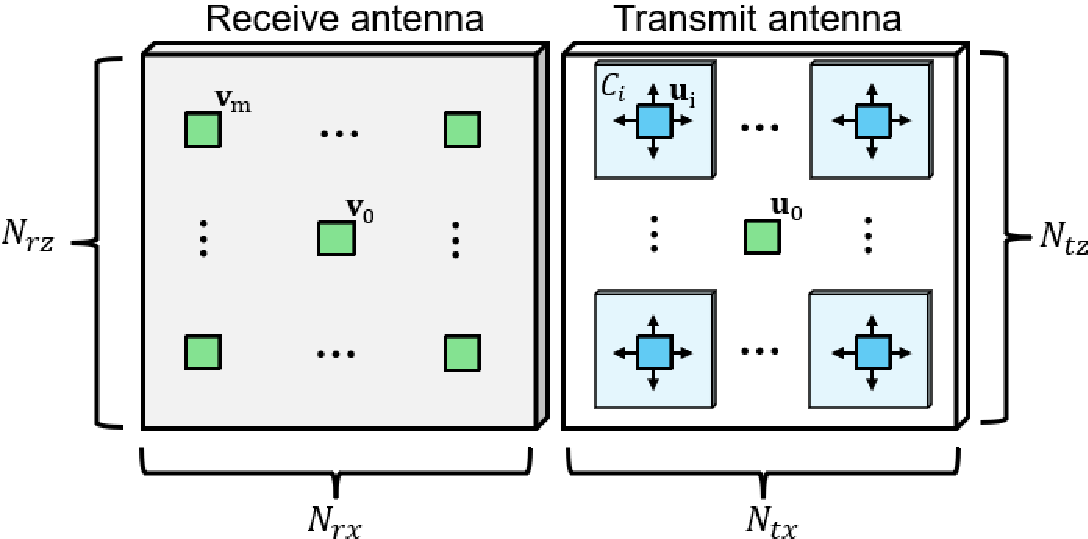}
    \caption{The transmitter with fluid antennas and receiver with fixed position antennas. The centroid of the transmit antennas is a fixed position antenna, denoted by $\mathbf{u}_0$.}
    \label{FA}
\end{figure}

We investigate an FA-enabled NF-ISCSC system with a single BS, where the transmit and receive antennas are co-located to facilitate both target detection and downlink semantic communication. In this communication-centric design, the BS is equipped with a planar array of $N_{\text{tx}} \times N_{\text{tz}}$ FAs for signal transmission, while $N_{\text{rx}} \times N_{\text{rz}}$ FPAs are used for signal reception. Each FA, indexed by $i \in \{0, \dots, N_{\text{tx}} \times N_{\text{tz}}-1\}$, is associated with a spatial coordinate $\mathbf{u}_i = [x_i, z_i]$ and is capable of the dynamic movement within a defined rectangular region $\mathcal{C}_i = [x_i^{\text{min}}, x_i^{\text{max}}]^T \times [z_i^{\text{min}}, z_i^{\text{max}}]$. The integration of FAs is intended to enhance downlink channel quality by providing spatial adaptability. In contrast, each FPA, indexed by $m \in \{0, \dots, N_{\text{rx}} \times N_{\text{rz}}-1\}$, is fixed at a specific coordinate $\mathbf{v}_m = [x_m, z_m]$. The antenna arrangement is illustrated in Fig. \ref{FA}. The centroid of the transmitter's FAs is assumed to be stationary, serving as the reference point for NF modelling. For the purpose of formulation, let the collection of coordinates of all FAs be denoted by $\mathbf{u} \in \mathbb{C}^{1 \times \left(2 \times N_{\text{tx}} \times N_{\text{tz}}\right)}$, such that
\begin{equation*}
    \begin{aligned}
       \mathbf{u} &= \Vect\left( \left[\mathbf{u}_0^T, ..., \mathbf{u}_i^T, ..., \mathbf{u}^T_{N_{\text{tx}} \times N_{\text{tz}}-1} \right] \right) \\
       & = \left[x_0, ..., x_i, ..., x_{N_{\text{tx}} \times N_{\text{tz}}-1}, z_0, ..., z_i, ..., z_{N_{\text{tx}} \times N_{\text{tz}}-1} \right],
    \end{aligned}
\end{equation*}
with $[\mathbf{x}_t,\;\mathbf{z}_t]$ collects the non-repeating $x$ and $z$ coordinates of all FAs. Similarly, for the coordinates of all FPAs, let $\mathbf{x}_r$ and $\mathbf{z}_r$ respectively collect the $x$ and $z$ coordinates. Then, we define the collection of FPA coordinates $\mathbf{v} \in \mathbb{C}^{1 \times \left(2 \times N_{\text{rx}} \times N_{\text{rz}}\right)}$ as
\begin{equation*}
    \begin{aligned}
        \mathbf{v} &= \Vect \left( \left[\mathbf{v}^T_0, ..., \mathbf{v}^T_m, ..., \mathbf{v}^T_{N_{\text{rx}} \times N_{\text{rz}}-1} \right]\right) \\
        & = \left[x_0, ..., x_m, ..., x_{N_{\text{rx}} \times N_{\text{rz}}-1}, z_0, ..., z_m, ..., z_{N_{\text{rx}} \times N_{\text{rz}}-1} \right],
    \end{aligned}
\end{equation*}
and the non-repeating $x$ and $z$ coordinates of all FPAs are collected by $[\mathbf{x}_r,\;\mathbf{z}_r]$.

The BS communicates with $K$ communication users (CUs), and each CU $k \in \mathcal{K}$ is equipped with a single antenna. To ensure the feasibility of beamforming design in a multi-user communication scenario, we set $K \leq (N_{\text{tx}} \times N_{\text{tz}})$. Simultaneously, the BS actively detects $L$ targets, each target $l \in \mathcal{L}$ being considered as an extended target. Alternative sensing models, such as the parametric scatterer model proposed in~\cite{liu2024crb}, represent extended targets using a reduced set of key scatterers and offer promising potential for efficient target representation. By capturing the essential scattering structure with fewer parameters, these models enhance computational efficiency without significantly sacrificing accuracy. In this work, we adopt the extended target model as a general framework, rather than focusing on specific parametric instances, to ensure broad applicability in system design. Nevertheless, our approach is inherently flexible and can be readily adapted to parametric models by reducing the number of scatterers per target to match the key scatterers required by a given model.

\begin{figure}[!t]
    \centering
    \includegraphics[width=0.8\linewidth]{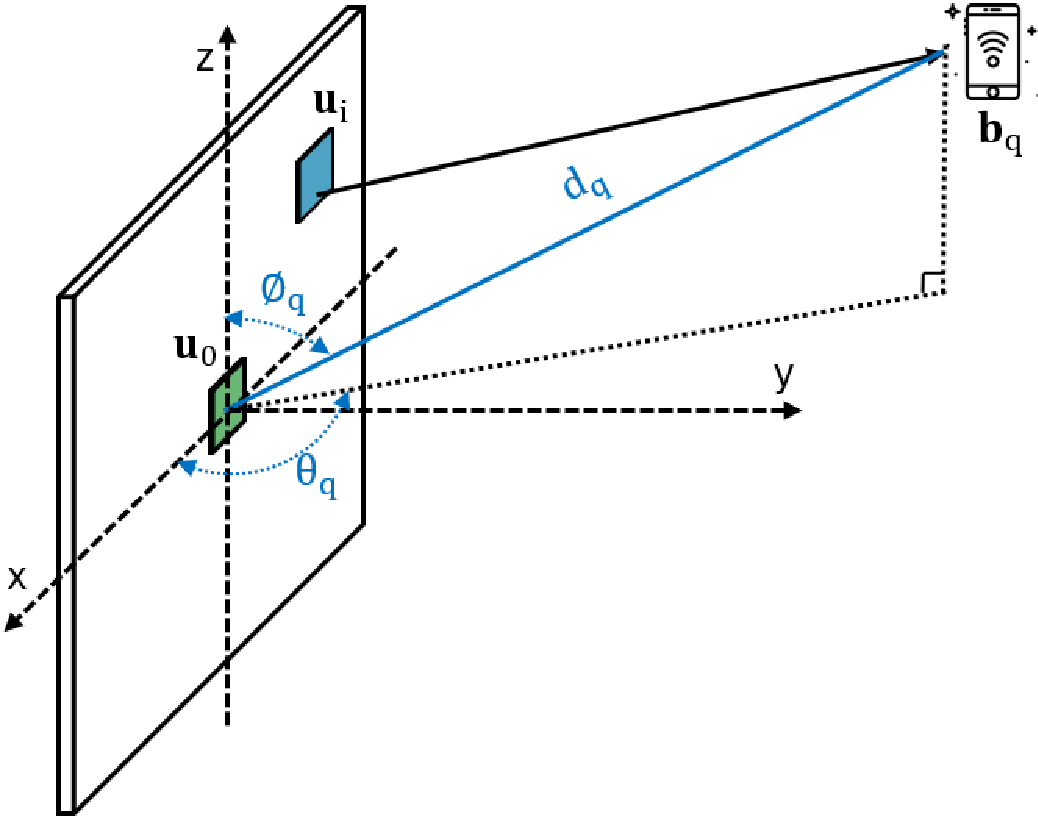}
    \caption{Spatial coordination of an object from the FAs.}
    \label{FA geometry}
\end{figure}

For multi-target sensing, we impose the condition $L \leq (N_{\text{tx}} \times N_{\text{tz}}) < (N_{\text{rx}} \times N_{\text{rz}})$ to ensure sufficient DoF for extracting information from the echo. Hence, we assume $(K + L) \leq (N_{\text{tx}} \times N_{\text{tz}}) < (N_{\text{rx}} \times N_{\text{rz}})$ throughout this paper. The location of the $k$-th CU or the $l$-th target is denoted by
\begin{equation*}
\begin{aligned}
        \mathbf{b}_q = [d_q \cos\left( \theta_q \right) \sin\left( \phi_q \right), & d_q \sin\left( \theta_q \right) \sin\left( \phi_q \right),\\
        & d_q \cos\left( \phi_q \right)], q \in [k,l],
\end{aligned}
\end{equation*}
where $\theta_q$, $\phi_q$, and $d_q$ represent the azimuth angle, the broadside angle, and the distance from the reference point of the transmit array (i.e., $\mathbf{u}_0$) to the object $q$, respectively. The geometric relationship is illustrated in Fig. \ref{FA geometry}.

\subsection{Signal Transmission Model}

\begin{figure*}[!t]
    \centering
    \includegraphics[width=0.8\linewidth]{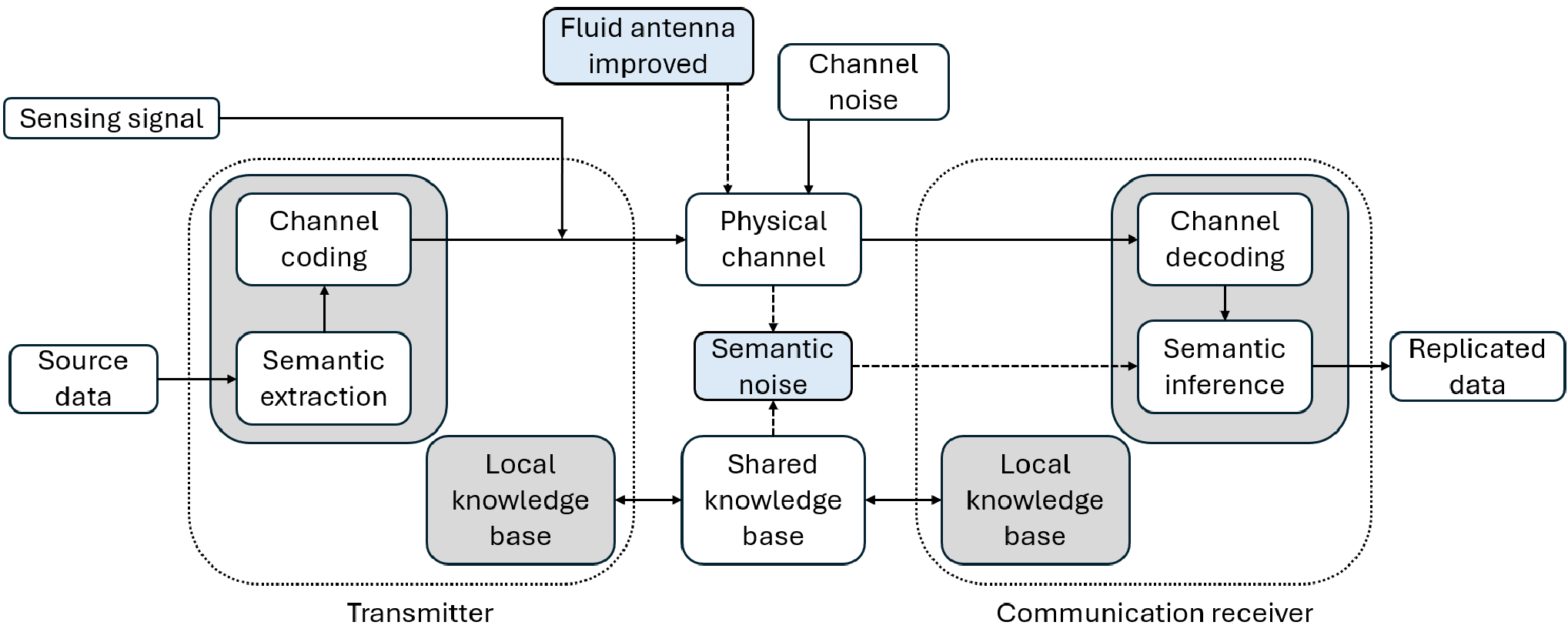}
    \caption{Semantic communication flow in the FA-enabled NF-ISCSC system.}
    \label{semcom blocks}
\end{figure*}

\begin{figure}[!t]
    \centering
    \includegraphics[width=0.9\linewidth]{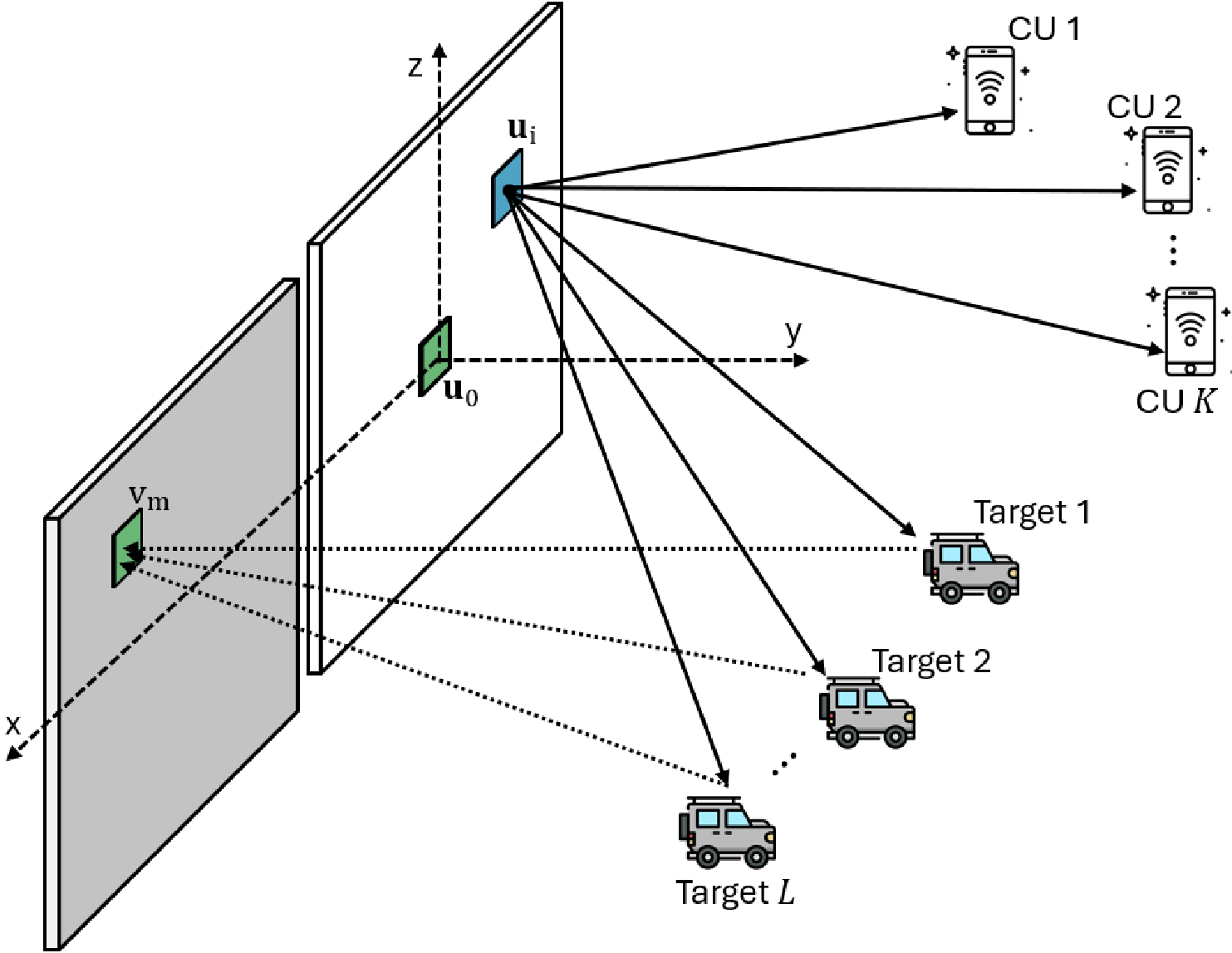}
    \caption{The proposed NF-ISCSC system with FAs.}
    \label{sys model}
\end{figure}

The semantic communication flow in the proposed framework is shown in Fig.~\ref{semcom blocks}. In semantic communication, semantic extraction is primarily guided by the local KBs of the sender and receiver. These local KBs are constructed from a shared KB and enriched with individual knowledge unique to each party. For example, in a vehicular network, the shared KB could be a global map, while each vehicle's local KB includes this global map along with specific local routes the vehicle has previously traversed. Based on the local KB, the sender focuses on extracting and conveying the underlying meaning of a message. To achieve this, a machine learning-based semantic extraction module is employed to identify key elements that capture the core semantic features of the message. For text-based messages, this typically involves leveraging models such as Transformers to select keywords that maximise performance metrics, such as the similarity between the original message and the reconstructed message at the receiver. The selected words are then encoded using channel coding techniques to ensure robustness against channel noise and transmission errors. A more integrated approach in semantic communication is to employ joint source-channel coding (JSCC), which is a machine learning-based approach such as proposed in \cite{xie2021deep, 10841377}, where the transmitted signal $\mathbf{C}$ is generated through a JSCC module $\mathcal{F}(\mathbf{S})$, with $\mathbf{S}$ denoting the source message. Therefore, for each user $k$, the semantic compression ratio is given by 
\begin{equation}\label{rho_k expression}
    \rho_k = \frac{\text{size} \left( \mathbf{c}_k \right)}{\text{size} \left( \mathbf{s}_k \right)},
\end{equation}
where $\mathbf{c}_k \in \mathbf{C}$ and $\mathbf{s}_k \in \mathbf{S}$.

At the receiver’s end, the local KB enables accurate reconstruction of the sender’s intended meaning, even if the exact transmitted words are not received. It is important to note that semantic noise plays a critical role in influencing the performance of semantic inference. Semantic noise may arise from poor channel conditions that cannot be fully mitigated by channel coding and decoding techniques, resulting in semantic distortion. It may also stem from impaired or incomplete KBs, which can lead to misinterpretation of the message \cite{cheng2024knowledge}. To address the former, the proposed framework employs FAs to enhance channel conditions and reduce the impact of transmission-related semantic noise. Semantic noise caused by impaired KBs is beyond the scope of this work, and we assume that the KBs are properly constructed and maintained.

As shown in Fig. \ref{sys model}, in the proposed framework, the BS simultaneously transmits semantic signals to the CUs and sensing signals to the targets through the utilisation of beamforming. The transmitted signal $\mathbf{X} \in \mathbb{C}^{(N_{\text{tx}} \times N_{\text{tz}}) \times F}$, which represents the joint communication and sensing signals with $F > N_{\text{tx}} \times N_{\text{tz}}$ frames, can be expressed as
\begin{equation}\label{ISCSC signal} 
\mathbf{X} = \mathbf{W}_c \mathbf{C}^H + \mathbf{W}_r, 
\end{equation} 
where $\mathbf{W}_c \in \mathbb{C}^{(N_{\text{tx}} \times N_{\text{tz}}) \times K}$ denote the communication beamforming matrix for the CUs, and each column corresponds to the beamforming vector for an individual user. Specifically, $\mathbf{w}_k \in \mathbb{C}^{(N_{\text{tx}} \times N_{\text{tz}}) \times 1}$ represents the beamforming vector for the $k$-th user. Thus, the complete beamforming matrix can be expressed as $\mathbf{W}_c = [\mathbf{w}_0, \mathbf{w}_1, \cdots, \mathbf{w}_{K-1}]$. The semantic communication signals intended for the CUs are denoted by $\mathbf{C} \in \mathbb{C}^{F \times K}$, where each column represents the signal intended for a specific user across $F$ frames. Specifically, $\mathbf{c}_k \in \mathbb{C}^{F \times 1}$ denotes the communication signal for the $k$-th CU. Accordingly, the communication signal matrix is given by $\mathbf{C} = [\mathbf{c}_0, \mathbf{c}_1, \cdots, \mathbf{c}_{K-1}]$. Additionally, $\mathbf{W}_r \in \mathbb{C}^{(N_{\text{tx}} \times N_{\text{tz}}) \times F}$ represents the sensing signal. Without loss of generality, if the number of frames $F$ is sufficiently large, we can make the following two assumptions \cite{liuxiang2020joint}:
\begin{enumerate}
    \item There is no correlation between the communication messages and the radar signal, i.e., $ \mathbb{E} \left[ \mathbf{C}^H \mathbf{W}_r^H \right] \approx \frac{1}{F} \mathbf{C}^H \mathbf{W}_r^H = \mathbf{0}_{K \times (N_{\text{tx}} \times N_{\text{tz}})}$.
    \item There is no correlation between the communication messages for different CUs, that is, $\mathbb{E} \left[ \mathbf{C}^H \mathbf{C} \right] \approx \frac{1}{F}\mathbf{C}^H \mathbf{C} = \mathbf{I}_{K}$.
\end{enumerate}

Hence, the covariance matrix of the transmitted signal $\mathbf{X}$ is given by
\begin{equation} \label{covar}
\begin{aligned}
        \mathbf{R}_x &= \mathbb{E} \left[ \mathbf{X} \mathbf{X}^H \right] \approx \frac{1}{F}\mathbf{X}\mathbf{X}^H \\
        & = \mathbf{W}_c \mathbf{W}_c^H + \frac{1}{F} \mathbf{W}_r \mathbf{W}_r^H = \sum_{k \in K} \mathbf{w}_k \mathbf{w}_k^H + \mathbf{R},
\end{aligned}
\end{equation}
where $\mathbf{R} \in \mathbb{C}^{(N_{\text{tx}} \times N_{\text{tz}}) \times (N_{\text{tx}} \times N_{\text{tz}})}$. The approximation is accurate when $F$ is sufficiently large.

\subsection{Communication Model}

After the BS transmits the signal $\mathbf{X}$, the received signal at the $k$-th CU can be expressed as: 
\begin{equation}\label{user received} 
\begin{aligned}
        &\mathbf{y}_k = \mathbf{\Phi}_k \otimes \mathbf{a}_t \left(\theta_k, \phi_k, d_k, \mathbf{u} \right) \mathbf{X} + \mathbf{n}_k = \mathbf{h}_k \mathbf{X} + \mathbf{n}_k \\
        &= \underbrace{\mathbf{h}_k \mathbf{w}_k \mathbf{c}^H_k}_{\text{Desired signal}} + \underbrace{ \mathbf{h}_{k} \sum_{k' \in K, k' \neq K} \mathbf{w}_{k'} \mathbf{c}^H_{k'}}_{\text{Multi-user interference}} + \underbrace{\mathbf{h}_k \mathbf{W}_r}_{\text{Sensing interference}} + \mathbf{n}_k,
\end{aligned}
\end{equation} 
where $\mathbf{\Phi}_k \in \mathbb{C}^{1 \times (N_{\text{tx}} \times N_{\text{tz}})}$ represents the free space path-loss vector for the $k$-th user, where each element is give by $\frac{1}{\sqrt{4 \pi || \mathbf{u}_i - \mathbf{b}_k||^2}}$ under a non-uniform spherical wave (NUSW) model \cite{liu2023near}, and $\mathbf{a}_t\left(\theta_k, \phi_k, d_k, \mathbf{u} \right) \in\mathbb{C}^{1 \times (N_{\text{tx}} \times N_{\text{tz}})}$ denotes the steering vector. Additionally, $\mathbf{n}_k \sim \mathcal{CN}(0,\sigma^2_c \mathbf{I}_{1 \times F})$ represents the communication noise.

On the target side, the received signal can be modelled as: 
\begin{equation}\label{eve receive} 
\begin{aligned}
    \mathbf{y}_l &= \sum_{s=1}^{N_s} \mathbf{\Phi}_{l,s} \otimes \mathbf{a}_t \left(\theta_{l,s}, \phi_{l,s}, d_{l,s}, \mathbf{u} \right) \mathbf{X} + \mathbf{n}_l \\
    &= \mathbf{h}_l \mathbf{X} + \mathbf{n}_l = \mathbf{h}_{l}
    \sum_{k \in K} \mathbf{w}_{k} \mathbf{c}_{k}^H + \mathbf{h}_l \mathbf{W}_r + \mathbf{n}_l,
\end{aligned}
\end{equation} 
where $N_s$ is the number of scatterers forming an extended target, and the subscript $s$ indicates the $s$-th scatterer. Additionally, $\mathbf{\Phi}_{l,s} \in \mathbb{C}^{1 \times (N_{\text{tx}} \times N_{\text{tz}})}$ is the path-loss vector where each element is given by $\frac{1}{\sqrt{4 \pi || \mathbf{u}_i - \mathbf{b}_{l,s}||^2}}$, and $\mathbf{a}_t\left(\theta_{l,s}, \phi_{l,s}, d_{l,s}, \mathbf{u} \right) \in \mathbb{C}^{1 \times (N_{\text{tx}} \times N_{\text{tz}})}$ denotes the corresponding steering vector for the target. The term $\mathbf{n}_l \sim \mathcal{CN} \left(0, \sigma^2_c \mathbf{I}_{1 \times F} \right)$ represents the noise encountered by the target. The received signal by target $l$ related to user $k$ is given by:
\begin{equation}
    \mathbf{y}_{l|k} = \underbrace{\mathbf{h}_{l} \mathbf{w}_{k} \mathbf{c}_{k}^H}_{\text{Intercepted signal}} + \underbrace{ \mathbf{h}_{l} \sum_{k' \in K, k' \neq k} \mathbf{w}_{k'} \mathbf{c}_{k'}^H + \mathbf{h}_l \mathbf{W}_r}_{\text{Interference}} + \mathbf{n}_l.
\end{equation}

Given that both the CUs and targets are located within the NF region of the BS, the steering vector can be characterised by \cite{liu2023near}:
\begin{equation}\label{eq1}
\footnotesize
        \mathbf{a}_t \left(\theta_q, \phi_q, d_q, \mathbf{u}\right) = \mathbf{a}_{\text{tx}}\left(\theta_q, \phi_q, d_q, \mathbf{x}_t \right) \otimes \mathbf{a}_{\text{tz}}\left(\phi_q, d_q, \mathbf{z}_t \right), q \in [k,l],
\end{equation}
where $\mathbf{a}_{\text{tx}}(\theta_q, \phi_q, d_q, \mathbf{x}_t) \in \mathbb{C}^{1 \times N_{\text{tx}}}$ and $\mathbf{a}_{\text{tz}}(\phi_q, d_q, \mathbf{z}_t) \in \mathbb{C}^{1 \times N_{\text{tz}}}$. 
Each element in the steering vector can be calculated using the following equations:
\begin{equation}\label{eq1.2}
\begin{aligned}
    & a_{\text{tx}}\left(\theta_q, \phi_q, d_q, x_i \right) \\
    &= e^{j\frac{2\pi}{\lambda} \left(x_i \cos\left(\theta_q\right) \sin\left(\phi_q\right) - \frac{x_i^2 \left( 1 - \cos^2\left(\theta_q\right) \sin^2\left(\phi_q\right)\right)}{2d_q}\right)}, \quad x_i \in \mathbf{x}_t, 
\end{aligned}
\end{equation}
and
\begin{equation}\label{eq1.3}
    a_{\text{tz}}\left(\phi_q, d_q, z_i \right) = e^{j\frac{2\pi}{\lambda} \left(z_i \cos\left(\phi_q \right)- \frac{z_i^2 \sin^2\left(\phi_q \right)}{2d_q}\right)}, \quad z_i \in \mathbf{z}_t.
\end{equation}

\subsection{Sensing Model}

After the BS transmits a joint signal to the locations of interest, the targets will reflect echo signals. The echo signal received by the BS, which contains information from all the targets, can be expressed as:
\begin{equation}\label{echo}
\footnotesize
\begin{aligned}
       \mathbf{Z} &= \sum_{l \in \mathcal{L}} \sum_{s=1}^{N_s} \mathbf{\Psi}_{l,s} \otimes \mathbf{a}_r^H \left(\theta_{l,s}, \phi_{l,s}, d_{l,s}, \mathbf{v} \right) \mathbf{a}_t \left(\theta_{l,s}, \phi_{l,s}, d_{l,s}, \mathbf{u} \right) \mathbf{X} + \mathbf{N} \\
       &= \mathbf{G} \mathbf{X} + \mathbf{N},
\end{aligned}
\end{equation}
where $\mathbf{N} \sim \mathcal{CN}\left(0, \sigma^2_{r} \mathbf{I}_{(N_{\text{rx}} \times N_{\text{rz}}) \times F}\right)$ represents the Gaussian noise, $\mathbf{\Psi}_{l,s} \in \mathbb{C}^{(N_{\text{rx}} \times N_{\text{rz}}) \times(N_{\text{tx}} \times N_{\text{tz}})}$ represents the path loss matrix with each element being $\frac{1}{4 \pi || \mathbf{u}_i - \mathbf{b}_{l,s}|| || \mathbf{v}_m - \mathbf{b}_{l,s}||}$, and $\mathbf{a}_r(\theta_{l,s}, \phi_{l,s}, d_{l,s}, \mathbf{v}) \in \mathbb{C}^{(1 \times N_{\text{rx}} \times N_{\text{rz}})}$ is the receiver steering vector, whose formulation is given below
\begin{equation}\label{eq2}
\begin{aligned}
    &\mathbf{a}_r\left(\theta_{l,s}, \phi_{l,s}, d_{l,s}, \mathbf{v} \right) \\
    & = \mathbf{a}_{\text{rx}}\left(\theta_{l,s}, \phi_{l,s}, d_{l,s}, \mathbf{x}_r \right) \otimes \mathbf{a}_{\text{rz}}\left(\phi_{l,s}, d_{l,s}, \mathbf{z}_r \right),
\end{aligned}
\end{equation}
where
\footnotesize
\begin{equation}\label{eq2.2}
\begin{aligned}
    &a_{\text{rx}}\left(\theta_{l,s}, \phi_{l,s}, d_{l,s}, x_m \right) = \\
    &e^{j\frac{2\pi}{\lambda} \left(m d_x \cos\left(\theta_{l,s} \right) \sin\left(\phi_{l,s} \right) - \frac{m^2 d_x^2 \left( 1 - \cos^2\left(\theta_{l,s} \right) \sin^2\left(\phi_{l,s} \right)\right)}{2 d_{l,s} }\right)}, x_m \in \mathbf{x}_r,
\end{aligned}
\end{equation}
\normalsize
and
\begin{equation}\label{eq2.3}
\begin{aligned}
    &a_{\text{rz}}\left(\phi_{l,s}, d_{l,s}, z_m \right) \\
    &= e^{j\frac{2\pi}{\lambda} \left(m d_z \cos\left(\phi_{l,s} \right)- \frac{m^2 d_z^2 \sin^2\left(\phi_{l,s} \right)}{2 d_{l,s}}\right)}, z_m \in \mathbf{z}_r.
\end{aligned}
\end{equation}

\section{Performance Measures}

In this section, we define various performance metrics for semantic communication, sensing, and computing. These metrics are essential for evaluating the quality of the communication, the precision and reliability of the sensing data, and the computational resources and power required for both tasks.

\subsection{Sensing} 

The mean square error (MSE) between the true and estimated values is a widely used metric for evaluating sensing performance. However, obtaining a closed-form expression for the MSE is often complex and computationally intensive, as noted in \cite{bekkerman2006target}. To address this challenge, we utilise the Cramér-Rao bound (CRB), which provides a theoretical lower bound on the variance of any unbiased estimator. The CRB serves as a benchmark for the minimum achievable MSE, defining the best possible estimation accuracy for a system under ideal conditions.

In the case of an extended target, the BS lacks prior knowledge of the number of scatterers or their respective angles and distances, as echoes are subject to random reflections. 
Instead of estimating the angles and distances of each individual scatterer, we estimate the entire echo channel matrix $\mathbf{G}$. Accurate estimation of the channel matrix enables the use of algorithms for extracting relevant information. For example, once the channel matrix is accurately estimated, the MUSIC algorithm can be applied to extract the angle information of each scatterer \cite{li1996adaptive, schmidt1986multiple}.

To compute the CRB, we first need to calculate the Fisher information matrix (FIM), which quantifies the amount of information contained in an observed variable about the unknown parameters of interest. By vectorizing $\mathbf{Z}$ and denoting it as $\mathbf{\bar{z}}$, we obtain:
\begin{equation}\label{linear model}
    \mathbf{\bar{z}}= \Vect \left({\mathbf{Z}} \right) = \left( \mathbf{X}^T \otimes \mathbf{I}_{\left(N_{\text{rx}} \times N_{\text{rz}}\right)} \right) \mathbf{\bar{g}} + \mathbf{\bar{n}},
\end{equation}
where $\mathbf{\bar{g}} = \Vect \left( \mathbf{G}\right)$ and $\mathbf{\bar{n}} = \Vect \left( \mathbf{N}\right)$. According to \cite{liu2021cramer,ben2009constrained}, the FIM of $\mathbf{\bar{g}}$ is given by
\begin{equation}\label{fim}
 \mathbf{J} = \frac{1}{\sigma^2_r} \mathbf{X}^* \mathbf{X}^T \otimes \mathbf{I}_{\left( N_{\text{rx}} \times N_{\text{rz}} \right) } = \frac{F}{\sigma_r^2} \mathbf{R}_x^T \otimes \mathbf{I}_{\left( N_{\text{rx}} \times N_{\text{rz}} \right)}.
\end{equation}

\begin{proof}
    See Appendix~\ref{FA TX RX}.
\end{proof}

In \eqref{fim}, the rank of the matrix \(\mathbf{X}\) is \(N_{\text{tx}} \times N_{\text{tz}}\), which is sufficient to fully recover the channel matrix \(\mathbf{G}\), whose rank is also \(N_{\text{tx}} \times N_{\text{tz}}\). As emphasised in \cite{ben2009constrained, liu2021cramer}, a sufficient rank ensures that the FIM remains non-singular. Hence, the CRB of $\mathbf{\bar{g}}$ is given by 
\begin{equation}
    \text{CRB}\left(\mathbf{G} \right) = \mathbf{J}^{-1} = \frac{\sigma_r^2 N_{\text{rx}} N_{\text{rz}}}{F} \Tr\left( \mathbf{R}_x^{-1}\right).
\end{equation}

\begin{remark}
Since the FIM of $\mathbf{\bar{g}}$ is independent of the locations of the FAs, the corresponding CRB is also unaffected. However, if the targets are modelled as point targets, as considered in \cite{ma2024movable, chen2025antenna}, both the FIM and the CRB become dependent on the positions of the FAs. Although a detailed system design of this aspect is beyond the scope of the present work, we provide the analytical derivation to show the dependence of FAs position on CRB for a point target in Appendix~\ref{point target} for completeness.
\end{remark}

\subsection{Semantic Communication}

According to \eqref{user received}, the signal-to-noise-plus-interference (SINR) ratio of the $k$-th CU is given by
\begin{equation}\label{cu sinr}
\begin{aligned}
       \gamma_k &= \frac{\mathbb{E} \left[\left| \left| \mathbf{h}_k \mathbf{w}_k \mathbf{c}^H_k \right| \right|^2 \right] }{ \sum_{k' \in \mathcal{K}, k' \neq k} \mathbb{E} \left[ \left| \left| \mathbf{h}_k \mathbf{w}_{k'} \mathbf{c}^H_{k'} \right| \right|^2\right] + \mathbb{E} \left[ \left| \left|\mathbf{h}_k \mathbf{W}_r \right| \right|^2 \right]+ \sigma_c^2} \\
       &= \frac{\left| \mathbf{h}_k \mathbf{w}_k \right|^2}{ \sum_{k' \in \mathcal{K}, k' \neq k} \left|\mathbf{h}_k \mathbf{w}_{k'} \right|^2 + \mathbf{h}_k \mathbf{R} \mathbf{h}_k^H + \sigma_c^2}.
\end{aligned}
\end{equation}

The semantic transmission rate is defined as the number of bits received by the user after decoding the semantic information from the received signal. The expression for the semantic transmission rate is given by \cite{yang2024secure}:
\begin{equation}\label{semantic rate}
    R_k = \frac{\iota}{\rho_k} \log_2 \left( 1 + \gamma_k \right),
\end{equation}
where the parameter $\rho_{LB} \leq \rho_{k} \leq 1$ represents the semantic extraction ratio shown in \eqref{rho_k expression}, and $\iota$ is a scalar value that converts the word-to-bit ratio. Additionally, $\rho_{LB}$ is the lower bound of $\rho_k$, with the formula provided in \cite[Lemma 1]{yang2024secure}:
\begin{equation}\label{rho lb}
    \rho_{LB} = \frac{1}{1 - \ln \varrho + \sum_{g=1}^G w_{g,k} \log p_{g,k}}.
\end{equation}

In \eqref{rho lb}, $\varrho$ represents the global lower bound of all individual Bilingual Evaluation Understudy (BLEU) scores. Additionally, $w_{g,k}$ denotes the weight assigned to the $g$-grams, where $G$ is the total number of $g$-grams required to represent a sentence. The precision score $p_{g,k}$ is user-specific and quantifies the accuracy of the semantic content recovered by the $k$-th CU.

In the context of joint transmission of sensing and communication signals, it is crucial to ensure that the unintended targets receive only a minimal amount of communication signal to prevent potential data breaches. To quantify the extent of information intercepted by an unintended receiver, it is necessary to model the amount of information captured by target $l$ from the communication intended for user $k$. The semantic transmission rate of target $l$ is defined as:
\begin{equation}\label{eve rate}
    R_{l|k} = \frac{\iota}{\rho_{l|k}} \log_2 \left( 1 + \Gamma_{l|k} \right),
\end{equation}
where $\rho_{l|k}$ is its semantic extraction ratio, $\Gamma_{l|k}$ is the SINR of target $l$ related to CU $k$. And $\Gamma_{l|k}$ can be derived from \eqref{eve receive}, that is:
\begin{equation}\label{eve snr}
    \Gamma_{l | k} = \frac{\left| \mathbf{h}_l \mathbf{w}_k \right|^2}{ \sum_{k' \in \mathcal{K}, k' \neq k} \left|\mathbf{h}_l \mathbf{w}_{k'} \right|^2 + \mathbf{h}_l \mathbf{R} \mathbf{h}_l^H + \sigma_c^2}.
\end{equation}

In semantic communication, the privacy of messages is closely linked to the security of the KBs \cite{zhaohui2024secure}. While shared KBs are essential for enabling efficient semantic encoding and decoding, they also introduce potential privacy vulnerabilities, particularly if an unintended user gains access to a similar or identical KB, which may allow for partial or full reconstruction of the transmitted message \cite{agarwal2022gdpr, won2024resource, guo2024survey}. Consequently, ensuring both the confidentiality of the transmitted content and the secure management of KBs is vital. To characterise the associated risks under the worst-case scenario, where KBs are assumed to be exposed (i.e., in the absence of protections such as those mandated by GDPR), we define the worst-case semantic secrecy rate (SSR) as the difference in semantic transmission rates between the legitimate receiver and an unintended user, both assumed to have access to the same KB (i.e., $\rho_k = \rho_{l|k}$). Accordingly, the worst-case SSR for the $k$-th CU can be formulated by incorporating \eqref{semantic rate} and \eqref{eve rate}:
\begin{equation}
\begin{aligned}
       S_k & = \max_{l \in \mathcal{L}} \left( R_{k} - R_{l|k}, \; 0 \right) \\
       & = \max_{l \in \mathcal{L}} \left( \frac{\iota}{\rho_k} \log_2\left( \frac{1 + \gamma_k}{1 + \Gamma_{l|k}}\right), \; 0 \right).
\end{aligned}
\end{equation}

If the unintended user has limited access to the KBs (i.e., GDPR in place), it cannot achieve the same level of semantic extraction as the legitimate user. In this case, \( \rho_{l|k} > \rho_k \), which leads to $S_{k}^{\text{limited}} > S_k$, indicating that the SSR under limited KB access is higher than under the worst-case assumption. Therefore, optimising the worst-case SSR (\( S_k \)) effectively provides a conservative lower bound, ensuring the system remains robust even in adversarial environments.

To better quantify the severity of potential data breaches, we propose using information efficiency, which is defined as the ratio between the semantic secrecy rate and the transmission rate of user $k$. The formulation is given by: 
\begin{equation}
    E_{k} = \frac{S_k}{R_k},
\end{equation}
where $0 \leq E_k \leq 1$. When $E_k$ approaches 1, it indicates a low probability or severity of a data breach event.

\subsection{Computing}
Extracting semantic information from traditional messages largely relies on advanced machine learning techniques, which introduce significant computational overhead. Therefore, it is crucial to consider computational power as a key component of the overall transmission power budget. As outlined in \cite{yang2024secure}, the computational power consumption is modelled using a natural logarithmic function to capture the relationship between computational complexity and power requirements. The formulation is given by:
\begin{equation}\label{eq18}
    P^{\text{Comp}} =  -\nu \sum_{k \in \mathcal{K}} \ln\left(\rho_{k}\right),
\end{equation}
where $\nu$ is a coefficient that converts the magnitude to its corresponding power. On the other hand, the power consumption for both communication and sensing at the BS is given by:
\begin{equation}\label{eq19}
    P^{\text{C\&S}} = \Tr\left(\mathbf{R}_x\right).
    \end{equation}

Moreover, after acquiring the sensing data, analysing and processing these data is crucial for applications such as DTs. This process requires computational resources and power, which must be accounted for. Since the echo signals are received in the same time slot, we consider a parallel computing model, where all computation tasks are computed at the same time. This requires an optimal CPU resource assignment. The computational task associated with the $l$-th target is represented by the tuple $\left(U_{l}, Q_{l}\right)$, where $U_{l}$ denotes the size of the sensing data in bits, $Q_{l}$ is the required computational resource in terms of CPU cycles per bit. As demonstrated in various studies, such as \cite{do2022digital,lu2020communication}, the processing latency for such tasks can be expressed as:
\begin{equation}
    T_{l} =  \frac{U_{l} Q_{l}}{f_{l} - \Bar{f}_{l}},
\end{equation}
where $f_{l}$ is the CPU frequency allocated by the BS for executing tasks related to target $l$, and $\bar{f}_{l}$ is the estimated CPU frequency required for the task. 

The power consumption for executing such a task, based on the CPU frequency and workload, can be formulated as:
\begin{equation}
    P_{l}^{\text{process}} = \kappa \left( f_{l} \right)^3 Q_{l},
\end{equation}
where $\kappa$ denotes the energy-efficiency coefficient that is dependent on the CPU design.

\section{Joint Design of FA-enabled NF-ISCSC System}

In this section, we formulate the problem, outline the key objectives and constraints, and present the algorithms designed to solve it. 

\subsection{Problem Formulation}
The design objective is to maximise the worst-case semantic secrecy rate, which simultaneously achieves the goal of enhancing the data rate. The optimisation problem is formulated as follows:
\begin{subequations} \label{opt1}
\begin{align}
    \max_{\mathbf{w}_k, \mathbf{R}_x, \mathbf{u}_i, f_l, \rho_k} \quad &   \min_{k \in \mathcal{K}} \left( S_k \right) \label{opt1a}\\
    \text{s.t.} \hspace{1cm} & \text{CRB}\left( \mathbf{G} \right) \leq \xi,  \label{opt1b}\\
    & P^{\text{C\&S}} + P^{\text{Comp}}  + \sum_{l \in \mathcal{L}}P_{l}^{\text{process}} \leq P_t,  \label{opt1c}\\
    & \max_{l \in \mathcal{L}} T_l \leq T^{\text{max}},
\label{opt1d}\\
    & \sum_{l \in \mathcal{L}} f_{l} \leq F^{\text{max}}, \label{opt1e}\\
    & p_{LB} \leq \rho_{k} \leq 1,  \forall k, \label{opt1f} \\
    & \mathbf{u}_i \in C_i, \forall i, i \neq 0, \label{opt1g}\\
    & \mathbf{R}_x \succeq \sum_{k \in \mathcal{K}} \mathbf{w}_k \mathbf{w}_k^H, \; \mathbf{w}_k \mathbf{w}_k^H \succeq 0,\forall k, \label{opt1h} \\
    & \text{rank} \left( \mathbf{w}_k \mathbf{w}_k^H\right) = 1, \forall k. \label{opt1i}
\end{align}
\end{subequations}

The constraint in \eqref{opt1b} guarantees the sensing performance by limiting the maximum CRB value to a predefined threshold ($\xi$). The constraint in \eqref{opt1c} bounds the total power consumption, encompassing the power allocated for signal transmission, semantic extraction, and data processing, to stay within the maximum available transmission power. The latency constraint in \eqref{opt1d} ensures that the worst-case task processing time at the BS does not exceed the maximum allowable delay ($T^{\text{max}}$). Similarly, \eqref{opt1e} limits the total allocated CPU resource for computational tasks to remain within the available processing capacity. The semantic extraction ratio for each user, denoted as $\rho_k$, is constrained by \eqref{opt1f}, ensuring it remains between a lower bound, $p_{LB}$, and 1. Finally, the movement constraint in \eqref{opt1g} restricts the FA positions.

The constraints \eqref{opt1h} and \eqref{opt1i} facilitate beamforming design by enforcing single-stream transmission for each user, thereby eliminating the need for advanced signal processing techniques \cite{wu2013physical}. In the absence of this constraint, more complex transceiver architectures, such as those employing water-filling power allocation or interference cancellation, would be necessary, resulting in significantly higher computational complexity.

In the following section, we relax the rank-one constraint due to its non-convex nature. The rank-one solution can be recovered through Gaussian randomisation \cite{wang2014outage}. Specifically, under the Gaussian randomisation method, random vectors are generated by sampling from a Gaussian distribution characterised by the obtained high-rank matrix. Each random vector is then evaluated using the original objective function, and the one achieving the best performance is selected as an approximate rank-one solution. To overcome the non-convexity of the objective function in \eqref{opt1}, we propose an AO approach. Specifically, we decompose \eqref{opt1} into three sub-problems and iteratively optimise each one until convergence. In the subsequent subsections, we detail each sub-problem and outline the algorithm for solving each sub-problem.

\subsection{Joint Beamforming and Computation Optimisation}

With given initial FA positions $\mathbf{u}_i$ and the initial semantic extraction ratios $\rho_k$, we can reformulate the original problem with respect to variables $[\mathbf{w}_k, \mathbf{R}_x, f_l, \zeta]$ as follows:
\begin{subequations}\label{opt2}
\begin{align}
    \max_{\mathbf{w}_k, \mathbf{R}_x, f_l, \zeta} \quad &   \zeta \label{opt2a}\\
    \text{s.t.}  \hspace{1cm} &  S_k \geq \zeta, \forall k, \label{opt2b} \\
    & \eqref{opt1b}-\eqref{opt1e}, \eqref{opt1h}.
\end{align}
\end{subequations}

In \eqref{opt2}, the first constraint is non-concave and presents a challenge for optimisation. To address this issue, we first reformulate \eqref{opt2b} as:
\begin{equation}\label{log expand}
    \log_2\left(A_k\right) - \log_2\left(B_k\right) + \log_2\left(C_{l|k}\right) - \log_2\left(D_{l|k}\right) \geq \zeta,
\end{equation}
where
\begin{equation}\label{rate terms list}
    \begin{cases}
    A_{k} = \sum_{k \in \mathcal{K}} \left|\mathbf{h}_k \mathbf{w}_{k} \right|^2 \\
    \vspace{0.2cm}
    \hspace{2cm} + \mathbf{h}_k \left(\mathbf{R}_x - \sum_{k \in \mathcal{K}} \mathbf{w}_k \mathbf{w}_k^H \right) \mathbf{h}_k^H + \sigma_c^2,\\
    B_{k} = \sum_{k' \in \mathcal{K}, k' \neq k} \left|\mathbf{h}_k \mathbf{w}_{k'} \right|^2 \\
    \vspace{0.2cm}
    \hspace{2cm} + \mathbf{h}_k \left(\mathbf{R}_x - \sum_{k \in \mathcal{K}} \mathbf{w}_k \mathbf{w}_k^H \right) \mathbf{h}_k^H + \sigma_c^2,\\
    C_{l|k} = \sum_{k' \in \mathcal{K}, k' \neq k} \left|\mathbf{h}_l \mathbf{w}_{k'} \right|^2 \\
    \vspace{0.2cm}
    \hspace{2cm} + \mathbf{h}_l \left(\mathbf{R}_x - \sum_{k \in \mathcal{K}} \mathbf{w}_k \mathbf{w}_k^H \right) \mathbf{h}_l^H + \sigma_c^2,\\
    D_{l|k} =  \sum_{k \in \mathcal{K}} \left|\mathbf{h}_l \mathbf{w}_{k} \right|^2 \\
    \hspace{2cm} + \mathbf{h}_l \left(\mathbf{R}_x - \sum_{k \in \mathcal{K}} \mathbf{w}_k \mathbf{w}_k^H \right) \mathbf{h}_l^H + \sigma_c^2.
    \end{cases}
\end{equation}

The second and fourth terms in \eqref{log expand} remain non-convex. To handle this, we approximate these terms using a first-order Taylor expansion, resulting in the following expression:
\begin{equation}\label{log approxi}
\begin{cases}
    \log_2\left(B_k\right) = \log_2\left(B_{k,e}\right) + \frac{1}{B_{k, e} \ln(2)} \left(B_k - B_{k, e}\right),\\
    
    \log_2\left(D_{l|k}\right) = \log_2\left(D_{l|k,e} \right) + \frac{1}{D_{l|k, e} \ln(2)} \left(D_{l|k} - D_{l|k, e}\right),\\
\end{cases}
\end{equation}
where the subscript $e$ denotes the value of the corresponding variable at each epoch. By substituting \eqref{log expand} and \eqref{log approxi}, we obtain the expression in \eqref{log terms} to replace \eqref{opt2b}. Consequently, the optimisation problem in \eqref{opt2} becomes convex and can be efficiently solved using established optimisation tools such as CVX \cite{grant2014cvx}.

\begin{remark}
Since the expression in \eqref{log terms} serves as a lower bound, a locally optimal solution can be obtained that guarantees a lower bound for problem \eqref{opt2}. However, it should be noted that the original optimisation problem \eqref{opt2} is inherently non-convex, and therefore, global optimality cannot be guaranteed. The final solution is generally dependent on the initial values (e.g., $B_{k,0}$ and $D_{l|k,0}$), and the algorithm typically converges to a stationary point that satisfies the Karush-Kuhn-Tucker (KKT) conditions.
\end{remark}

\subsection{FA Position Optimisation}

For any given values of $\mathbf{w}_k$, $\mathbf{R}_x$, $f_l$ and $\rho_k$, the positions of the FAs $\mathbf{u}_i$ can be determined by solving the following optimisation problem:
\begin{subequations} \label{opt FA}
\begin{align}
    \max_{\mathbf{u}_i} \quad &  \min_{k \in \mathcal{K}} \left( S_k \right) \label{opt FA a}\\
    \text{s.t.} \quad &  \mathbf{u}_i \in C_i, \forall i, i \neq 0. \label{opt FA b}
\end{align}
\end{subequations}

The optimisation problem in \eqref{opt FA} is non-convex due to the non-convex nature of the objective function in \eqref{opt FA a}. To address this, we first describe a benchmark algorithm widely used in the literature, such as in \cite{ma2024multi, feng2024weighted, lyu2024flexible}. We then propose a new algorithm that reduces computational complexity while maintaining performance.

\subsubsection{Benchmark method (second-order Taylor expansion)} 

To apply this method, we first replace $\min_{k \in \mathcal{K}} \left( S_k \right)$ with the constraint $S_k \geq \xi, \forall k$, where $\xi$ is the auxiliary variable that we seek to maximise. To address the non-convexity of $S_k$, a second-order Taylor expansion is employed to construct a lower bound for $S_k$. Recall that $S_k$ can be expressed as in \eqref{log expand}, the lower bound of $S_k$ can be found using second-order Taylor expansion. The lower bound of $S_k$ is determined by finding the lower bound of $R_k$ and the upper bound of $R_{l|k}$. Specifically, to find the lower bound of $R_k$, we must derive the lower bound for $\log_2\left( A_k \right)$ and the upper bound for $\log_2\left( B_k \right)$, as shown in \eqref{log expansion term 1} and \eqref{log expansion term 2}, respectively. In \eqref{log expansion term 1} and \eqref{log expansion term 2}, the Jacobian and Hessian matrices are denoted by $\nabla \in \mathbb{C}^{1 \times \left(2 \times N_{\text{tx}} \times N_{\text{tz}}\right)}$ and $\nabla^2 \in \mathbb{C}^{\left(2 \times N_{\text{tx}} \times N_{\text{tz}}\right) \times \left(2 \times N_{\text{tx}} \times N_{\text{tz}}\right)}$, respectively. Additionally, the vector $\mathbf{u}$ has been defined in Section. II.

Likewise, the upper bound of $R_{l|k}$ is determined by finding the lower bound of $\log_2 \left( C_{l|k} \right)$ and the upper bound of $\log_2 \left( D_{l|k} \right)$, as shown in \eqref{log expansion term 3} and \eqref{log expansion term 4}, respectively. In this manner, the constraint in \eqref{opt FA b} is replaced by the objective in \eqref{log expansion obj}, with the subscript $e$ indicating the number of epochs. To ensure consistency in matrix dimensions, the Hessian matrices are replaced by scalar values $\epsilon$ and $\delta$. The selection of $\epsilon$ and $\delta$ in \eqref{log expansion term 1}–\eqref{log expansion term 4} should be made to avoid the appearance of $\mathbf{u} \leq \mathbf{u}_e$, which would result in a negative gradient that minimises the objective function instead of maximising it. A recommended approach is to ensure that the conditions \(\epsilon_k \mathbf{I} \succeq \nabla^2 \log_2 \left(A_{k,e}\right)\), \(\delta_k \mathbf{I} \succeq \nabla^2 \log_2 \left(B_{k,e}\right)\), \(\epsilon_{l|k} \mathbf{I} \succeq \nabla^2 \log_2 \left(C_{l|k, e}\right)\), and \(\delta_{l|k} \mathbf{I} \succeq \nabla^2 \log_2 \left(D_{l|k, e}\right)\) hold, as shown in \cite{sun2016majorization, boyd2004convex}. One practical choice for \(\epsilon\) and \(\delta\) is to use the maximum eigenvalues of the respective matrices, ensuring that the conditions are satisfied while also improving the stability of the optimisation process.

\begin{table*}
\footnotesize
\centering
\begin{minipage}{1\textwidth}
    \begin{align} 
        & g\left(\mathbf{w}_k, \mathbf{R}_x \right) \overset{\Delta}{=} \log_2\left(A_k\right) - \log_2\left(B_{k,e}\right) - \frac{1}{B_{k, e} \ln(2)} \left(B_k - B_{k,e}\right) + \log_2\left(C_{l|k}\right) - \log_2\left(D_{l|k,e}\right) - \frac{1}{D_{l|k,e} \ln(2)} \left(D_{l|k} - D_{l|k,e}\right), \label{log terms}\\
        &\log_2\left(A_k\right) \geq \log_2\left(A_{k,e}\right) + \nabla\log_2\left(A_{k,e}\right) \left( \mathbf{u} - \mathbf{u}_e \right)^T - \left( \mathbf{u} - \mathbf{u}_e \right) \frac{\nabla^2 \log_2 \left(A_{k,e}\right)}{2} \left( \mathbf{u} - \mathbf{u}_e \right)^T , \quad  \label{log expansion term 1}\\
        &\log_2\left(B_k\right) \leq \log_2\left(B_{k,e}\right) + \nabla\log_2\left(B_{k,e}\right) \left( \mathbf{u} - \mathbf{u}_e \right)^T + \left( \mathbf{u} - \mathbf{u}_e \right)\frac{\nabla^2 \log_2\left(B_{k,e}\right)}{2} \left( \mathbf{u} - \mathbf{u}_e \right)^T , \label{log expansion term 2}\\
        & \log_2\left(C_{l|k}\right) \geq \log_2\left(C_{l|k, e}\right) + \nabla\log_2\left(C_{l|k, e}\right) \left( \mathbf{u} - \mathbf{u}_e \right)^T - \left( \mathbf{u} - \mathbf{u}_e \right)\frac{\nabla^2 \log_2 \left(C_{l|k, e}\right)}{2} \left( \mathbf{u} - \mathbf{u}_e \right)^T , \label{log expansion term 3}\\
        & \log_2\left(D_{l|k}\right) \leq \log_2\left(D_{l|k, e}\right) + \nabla\log_2\left(D_{l|k, e}\right) \left( \mathbf{u} - \mathbf{u}_e \right)^T + \left( \mathbf{u} - \mathbf{u}_e \right)\frac{\nabla^2 \log_2 \left(D_{l|k, e}\right)}{2} \left( \mathbf{u} - \mathbf{u}_e \right)^T ,  \label{log expansion term 4}\\
        &  g\left( \mathbf{u} \right) \overset{\Delta}{=}  \log_2\left(A_{k, e}\right) + \nabla\log_2\left(A_{k, e} \right) \left( \mathbf{u} - \mathbf{u}_e \right)^T - \left( \mathbf{u} - \mathbf{u}_e \right) \frac{\nabla^2 \log_2 \left(A_{k, e}\right)}{2} \left( \mathbf{u} - \mathbf{u}_e \right)^T  -\log_2\left(B_{k, e}\right) - \nabla\log_2\left(B_{k, e}\right) \left( \mathbf{u} - \mathbf{u}_e \right)^T \nonumber \\
        &- \left( \mathbf{u} - \mathbf{u}_e \right)  \frac{\nabla^2 \log_2 \left(B_{k, e}\right)}{2} \left( \mathbf{u} - \mathbf{u}_e \right)^T + \log_2\left(C_{l|k, e}\right) + \nabla\log_2\left(C_{l|k, e}\right) \left( \mathbf{u} - \mathbf{u}_e \right)^T - \left( \mathbf{u} - \mathbf{u}_e \right) \frac{\nabla^2 \log_2 \left(C_{l|k, e}\right)}{2} \left( \mathbf{u} - \mathbf{u}_e \right)^T  \nonumber\\
        & -\log_2\left(D_{l|k, e}\right) - \nabla\log_2\left(D_{l|k, e}\right) \left( \mathbf{u} - \mathbf{u}_e \right)^T - \left( \mathbf{u} - \mathbf{u}_e \right) \frac{\nabla^2 \log_2 \left(D_{l|k, e}\right)}{2} \left( \mathbf{u} - \mathbf{u}_e \right)^T . \label{log expansion obj}
    \end{align}
\medskip
\hrule
\end{minipage}
\end{table*}

By substituting each Hessian matrix in \eqref{log expansion obj} with the corresponding scalar values \([\delta_k, \epsilon_k, \delta_{l|k}, \epsilon_{l|k}]\), the optimisation problem in \eqref{opt FA} becomes convex and can be efficiently solved using existing optimisation tools. The computational complexity at each epoch is \(\mathcal{O} \left( \left( 2 N_{\text{tx}} N_{\text{tz}} \right)^3 \right)\).

\subsubsection{Projected BFGS}

The high computational complexity of the benchmark method primarily stems from the calculation of the Hessian matrix at each iteration, which scales cubically with the size of FAs. As a result, when the number of FAs is large, the benchmark method becomes inefficient. To address this challenge, we propose a reduced-complexity approach based on the projected BFGS algorithm. This method combines the efficiency of the projected gradient descent with the approximation capabilities of the BFGS algorithm. The details of the projected BFGS algorithm are outlined below.

By differentiating the function \eqref{log expansion obj} with respect to the position vector $\mathbf{u}$ and setting the result equal to zero, we can formulate the gradient descent problem as follows:
\begin{equation}\label{pgd update}
       \mathbf{u}_{e+1} = \mathbf{u}_{e} + \tau \left(\nabla_e^2\right)^{-1} \nabla_e, 
\end{equation}
where $\tau$ is the step size, and $\left(\nabla_e^2\right)^{-1} \nabla_e$ represents the Newton step. The Newton step is used here because the second-order search direction provides faster convergence than a first-order direction, since it takes into account both the direction and the curvature of the objective function. Based on \eqref{log expansion obj}, the Newton step can be computed as follows:
\begin{equation}
    \begin{aligned}
        &\nabla_e = \nabla\log_2\left(A_{k, e}\right) - \nabla\log_2\left(B_{k, e}\right) \\
        \vspace{0.4cm}
        &\hspace{2cm} + \nabla\log_2\left(C_{l|k, e} \right) - \nabla\log_2\left(D_{l|k, e} \right),\\
        &\nabla_e^2 = \nabla^2 \log_2 \left(A_{k, e}\right) + \nabla^2 \log_2 \left(B_{k, e}\right) \\
        &\hspace{2cm} + \nabla^2 \log_2 \left(C_{l|k, e}\right) + \nabla^2 \log_2 \left(D_{l|k, e}\right).
    \end{aligned}
\end{equation}

To ensure that $\mathbf{u}_{i, e+1}$ satisfies the constraint in \eqref{opt FA b}, we employ a projection function $\mathcal{P}$ that projects $\mathbf{u}_i$ onto the feasible region. The projection function $\mathcal{P}$ is defined as follows:

\begin{equation}\label{projection func}
\begin{aligned}
    \mathcal{P}(\mathbf{u}_{i,e+1}) =
    \big[
        &\max\left(x_i^{LB},\ \min(x_{i,e+1},\ x_i^{UB})\right),\\
        &\max\left(z_i^{LB},\ \min(z_{i,e+1},\ z_i^{UB})\right)\big],
\end{aligned}
\end{equation}
where $x_i^{LB}$, $z_i^{LB}$, $x_i^{UB}$ and $z_i^{UB}$ are the lower bound and the upper bound of the x and z coordinates, respectively.

To avoid the computational overhead of calculating the Hessian matrix in every iteration, we adopt the BFGS method, which approximates the Hessian matrix at each step.
\begin{lemma}[BFGS \cite{fletcher2000practical}]\label{bfgs}
\begin{equation}
    \begin{aligned}
        \left(  \nabla_{e+1}^{2} \right)^{-1} &= \bigg( \mathbf{I} - \frac{\Delta \left( \nabla_{e+1}^T\right) \Delta \mathbf{u}_{e+1} }{\Delta \left( \nabla_{e+1} \right) \Delta \mathbf{u}_{e+1}^T }\bigg) \left(  \nabla_{e}^{2} \right)^{-1} \\
        & \; \times \bigg( \mathbf{I} - \frac{ \Delta \mathbf{u}_{e+1}^T \Delta \left( \nabla_{e+1} \right)}{\Delta \left( \nabla_{e+1} \right) \Delta \mathbf{u}_{e+1}^T }\bigg) + \frac{\Delta \mathbf{u}_{e+1}^T \Delta \mathbf{u}_{e+1}}{\Delta \left( \nabla_{e+1} \right) \Delta \mathbf{u}_{e+1}^T },    
    \end{aligned}
\end{equation}
where $\Delta \mathbf{u}_{e+1} = \mathbf{u}_{e+1} - \mathbf{u}_{e}$ and $\Delta \left( \nabla_{e+1} \right) = \nabla \left( \mathbf{u}_{e+1} \right) - \nabla \left( \mathbf{u}_{e} \right)$.
\end{lemma}

Therefore, according to \textit{\textbf{Lemma \ref{bfgs}}}, the Hessian matrix can be approximated by only computing the Jacobian matrix at each iteration. Additionally, the initial inverse Hessian matrix, denoted as $\left( \nabla^{2}_{0} \right) ^{-1}$, can be generated randomly.

Finally, the step size $\tau$ must be selected carefully to ensure convergence and stability. To determine an appropriate step size at each iteration, we employ the backtracking line search method. Backtracking line search is an adaptive step size selection technique commonly used in gradient-based optimisation algorithms. Starting with an initial step size, the method iteratively reduces it by a fixed factor until a sufficient change in the objective function is achieved, according to the Armijo condition. The per-step procedure for the projected BFGS algorithm with BLS is outlined in Algorithm~\ref{alg1}.

\begin{algorithm}
\caption{Projected BFGS algorithm with BLS}\label{alg1}
\begin{algorithmic}[1]
\STATE With given $\mathbf{w}_k$, $\mathbf{R}_x$, $\rho_k$, $f_l$, a initial value $\mathbf{u}_{i, 0}$ a initial value of step size $\tau$, and a initial value of $\left(\nabla_{0}^2\right)^{-1}$.
    \STATE Calculate the Newton step $\left(\nabla_e^2\right)^{-1} \nabla_e$.
    \vspace{0.2cm}
    \WHILE{$g\left(\mathbf{u}_{i, e+1} \right) < g\left(\mathbf{u}_{i, e} \right) + \tau \nabla_e \left(\nabla_e^2\right)^{-1} \nabla_e^T$}
            \STATE Apply the BLS to find $\tau$.
    \ENDWHILE
    \STATE Update $\mathbf{u}_{e+1}$ according to \eqref{pgd update}.
    \STATE Project $\mathbf{u}_{e+1}$ using \eqref{projection func}.
    \vspace{0.2cm}
    \STATE Update $\left(\nabla_{e+1}^2\right)^{-1}$ by using \textit{\textbf{Lemma \ref{bfgs}}}.
\end{algorithmic}
\end{algorithm}

The per epoch complexity of the projected BFGS algorithm is $\mathcal{O} \left( \left(2 N_{\text{tx}} N_{\text{tz}} \right)^2  +   \left(2 N_{\text{tx}} N_{\text{tz}} \right) \log\left(\frac{1}{\tau_{\min}} \right)   \right)$. Here, 
$\tau_{\min}$ is the smallest step size tolerated in the BLS algorithm. The first term in the complexity expression accounts for the cost of updating the Hessian approximation, while the second term reflects the iterative process of adjusting the step size, with the logarithmic factor corresponding to the number of iterations required to reduce the step size until it meets the minimum threshold.

\begin{remark}
    The objective function in \eqref{log expansion obj} represents a lower bound of $S_k$. The projected BFGS algorithm is employed to locally maximise this bound. Since the original optimisation problem \eqref{opt FA} is non-convex, the projected BFGS method converges to a stationary point, corresponding to a locally optimal lower bound. Although global optimality cannot be guaranteed, the use of a backtracking line search ensures a proper step size at each iteration, which helps guide the algorithm toward feasible and potentially tight solutions.
\end{remark}

\subsection{Semantic Extraction Ratio Optimisation}
For any given values of $\mathbf{w}_k$, $\mathbf{R}_x$, $f_l$ and $\mathbf{u}_i$, the optimal values of $\rho_k$ can be determined by solving the following optimisation problem:
\begin{subequations} \label{opt semantic}
\begin{align}
    \min_{\rho_k} \quad &   \sum_{k \in \mathcal{K}} \rho_k \label{opt semantic a}\\
    \text{s.t.} \quad & P^{\text{Comp}} \leq P_t - P^{\text{C\&S}} - \sum_{l \in \mathcal{L}} P_{l}^{\text{process}}, \label{opt semantic b}\\
    & p_{LB} \leq \rho_{k} \leq 1,  \forall k. \label{opt semantic c}
\end{align}
\end{subequations}

Optimisation problem \eqref{opt semantic} is convex and can be solved by using the bisection search method. As problem \eqref{opt semantic} is a box-constrained linear program, it admits a unique global optimum, which can be efficiently computed.

The overall procedure for solving the optimisation problem \eqref{opt1} is outlined in Algorithm \ref{alg3}. In this algorithm, the stopping criterion is defined as $\left|\min_{k \in \mathcal{K}} \left( S_{k, e+1} \right) - \min_{k \in \mathcal{K}} \left( S_{k, e} \right) \right| \leq \varsigma$, where $S_{k,e}$ denotes the semantic secrecy rate at the $e$-th iteration for user $k$, and $\varsigma$ is a small predefined threshold that determines the convergence condition. When the difference between the minimum semantic secrecy rates in consecutive iterations is less than or equal to $\varsigma$, the algorithm stops, indicating that the optimisation has converged. A convergence analysis is included in Appendix~\ref{coverge Appendix}.

\begin{algorithm}
\caption{Alternating optimisation algorithm}\label{alg3}
\begin{algorithmic}[1]
\REPEAT
    \STATE With given $\mathbf{u}_{e}$ and $\rho_k$, solve optimisation problem \eqref{opt2} with replacing \eqref{opt2b} with \eqref{log terms}.
    \STATE Solve optimisation problem \eqref{opt FA} by using Algorithm \ref{alg1}.
    \STATE Solve optimisation problem \eqref{opt semantic} by using the bisection search method.
    \STATE Update the epoch with $e = e+1$.
\UNTIL{Stopping criterion is satisfied}
\end{algorithmic}
\end{algorithm}

\begin{remark}
In the proposed alternating optimisation framework, sub-problems~1 and~2 are non-convex and are solved to local optima of their respective lower bounds, while sub-problem~3 is convex and admits a unique global solution. As a result, the overall procedure converges to a stationary point of the original problem. However, due to the non-convexity of the original problem, global optimality of the final solution cannot be guaranteed.
\end{remark}

\section{Simulation Results}

In this section, we present numerical results to evaluate the effectiveness of the proposed design. The default values for the parameters are summarised in Table \ref{table1}. The simulation utilises the mmWave frequency band, where the small antenna size leads to narrower movable regions for the FAs compared to existing works, such as \cite{qin2024cramer, feng2024weighted}. For simplicity, we assume that the movable region of the FAs is square-shaped. Consequently, the movable range along the x-axis or z-axis is $\sqrt{C_i}$. While the analysis can be extended to non-square movable regions, this falls beyond the scope of the current study.

\begin{table}[!t]
    \centering
    \caption{List of parameters.}
    \begin{tabular}{|c|c|}
        \hline
         Parameter & Value\\
         \hline
         \hline
         Carrier frequency & $f_c = 50$ GHz\\
         \hline
         Number of FAs & $N_{\text{tx}} = N_{\text{tz}} = 3$\\
         \hline
         Number of FPs & $N_{\text{rx}} = N_{\text{rz}} = 5$\\
         \hline
         Number of frames & $F = 100$\\
         \hline
         Movable region for each FA & $C_i = 0.0025 \text{m}^2$\\
         \hline
         Number of CUs & $K = 5$\\
         \hline
         Number of targets & $L = 2$\\
         \hline
         Number of scatter for each target & $N_s = 6$\\
         \hline
         Transmit power budget & $P_t = 25$ dBm\\
         \hline
         Communication noise power & $\sigma_c^2 = -30$ dBm\\
         \hline
         Sensing noise power & $\sigma_r^2 = -40$ dBm\\
         \hline
         CRB upper bound & $\xi = 0.5$\\
         \hline
         Maximum tolerance processing latency & $T^{\text{max}} = 0.02 s$\\
         \hline
         Sensing data size & $U_l = 0.5 $ Mb\\
         \hline
         Required computational resource & $Q_l = 1.1 \times 10^2$ cycles/bit \\
         \hline
         Energy-efficiency coefficient & $\kappa = 10^{-28}$\\
         \hline
    \end{tabular}
    \label{table1}
\end{table}

\subsection{Algorithm Performance}

\begin{figure}[!t]
    \centering
    \includegraphics[width=0.9\linewidth]{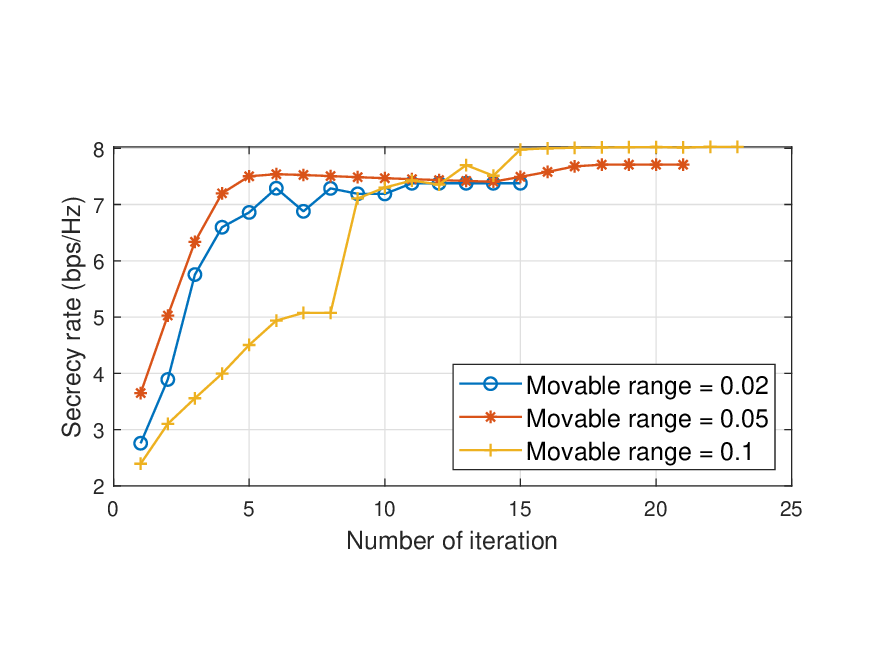}
    \caption{Convergence Performance of Algorithm \ref{alg1}.}
    \label{converge}
\end{figure}

\begin{figure}[!t]
    \centering
    \includegraphics[width=0.8\linewidth]{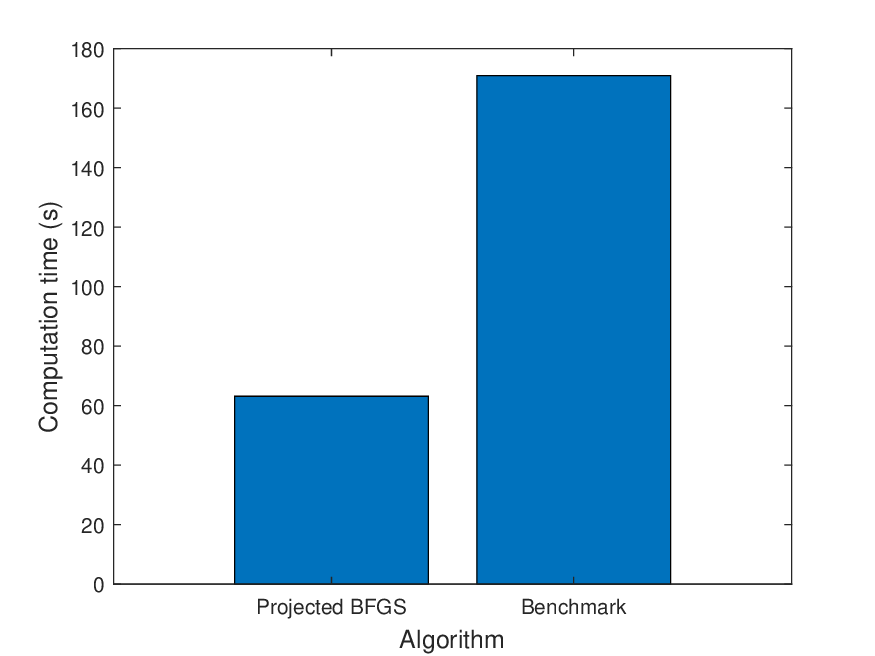}
    \caption{Computation time of Algorithm \ref{alg1} and the Benchmark algorithm.}
    \label{Comp time}
\end{figure}

Fig.~\ref{converge} illustrates the convergence performance of the secrecy rate with varying movable range parameters, specifically set to 0.02, 0.05, and 0.1. As shown in the figure, the secrecy rate rapidly increases in the initial iterations and stabilises after approximately 10 to 15 iterations across all three cases. Notably, a larger movable range generally results in slower convergence due to the expanded search space. The convergence behaviours indicate that beyond a certain number of iterations, minimal gains in secrecy rate are observed, demonstrating efficient convergence of the algorithm.

Fig.~\ref{Comp time} compares the computational efficiency of the proposed Projected BFGS algorithm with the Benchmark algorithm based on the second-order Taylor expansion. Computation times were recorded using an Intel\textsuperscript{\textregistered}{} Core\textsuperscript{TM}{} i7-12700H processor. The Projected BFGS algorithm significantly outperforms the Benchmark, reducing computation time by approximately 60\%. Specifically, the proposed method requires about 63 seconds per iteration, compared to 170 seconds for the Benchmark, demonstrating the superior computational efficiency of the proposed optimisation approach.

\subsection{Semantic Communication Performance}

\begin{figure}[!t]
    \centering
    \includegraphics[width=0.8\linewidth]{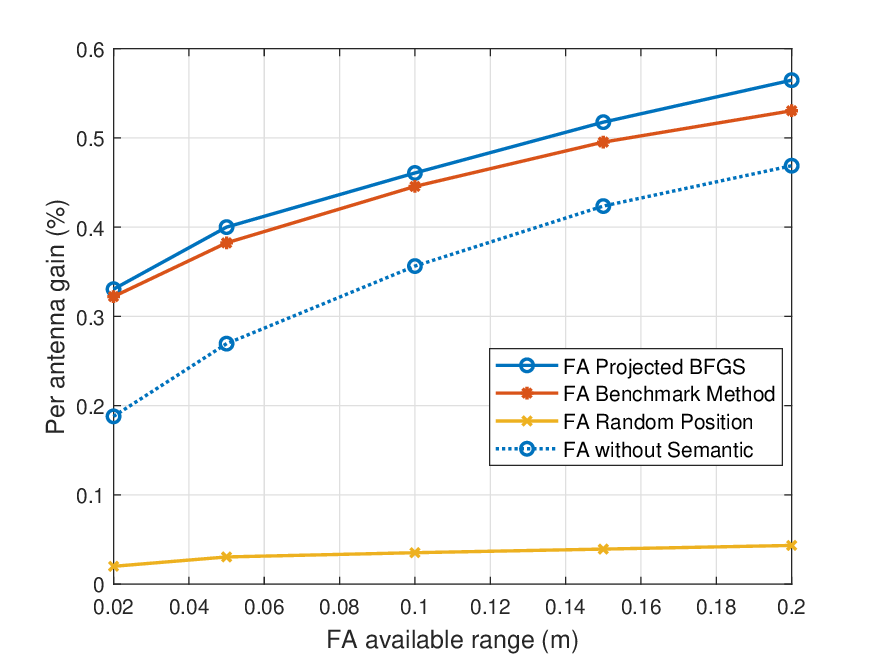}
    \caption{Per antenna gain against FAs movable range $\sqrt{C_i}$.}
    \label{fig gain vs ma}
\end{figure}

Fig.~\ref{fig gain vs ma} illustrates the relationship between the movable range of FAs, denoted as $\sqrt{C_i}$, and the resulting per antenna gain compared with the case of using FPAs, expressed as a percentage. Three methods are compared: the benchmark method, the projected BFGS method, and randomised FAs, where the positions of FAs are randomly generated within the movable range. The result reveals that both the benchmark method and the projected BFGS benefit from increasing the movable range $\sqrt{C_i}$, as larger movable regions allow for more exploration of better antenna positioning, leading to significant performance gains. For example, when the movable range is 20 cm, the per antenna gain for the benchmark method is 0.53\% while the projected BFGS gives 0.56\%. Notably, the projected BFGS consistently outperforms the benchmark method across all $\sqrt{C_i}$ values, and the difference in the performance becomes more notable as $\sqrt{C_i}$ increases. This demonstrates the superior optimisation capabilities of the projected BFGS algorithm, on top of its lower complexity when compared with the benchmark method. In contrast, the random method exhibits a nearly flat and negligible performance gain. For instance, with 20cm of movable range, the per antenna gain is only 0.04\%, which is ten times less than the other two algorithms can achieve. The figure also compares the projected BFGS algorithm with and without using semantic. Notably, incorporating semantic processing consistently enhances the per-antenna gain by approximately 1.2–1.5 times compared to the non-semantic design (i.e., conventional ISAC designs as in \cite{zhao2024modeling,hao2024fluid}). This clearly highlights the advantage of integrating semantic into the model.

\begin{figure}[!t]
    \centering
    \includegraphics[width=0.8\linewidth]{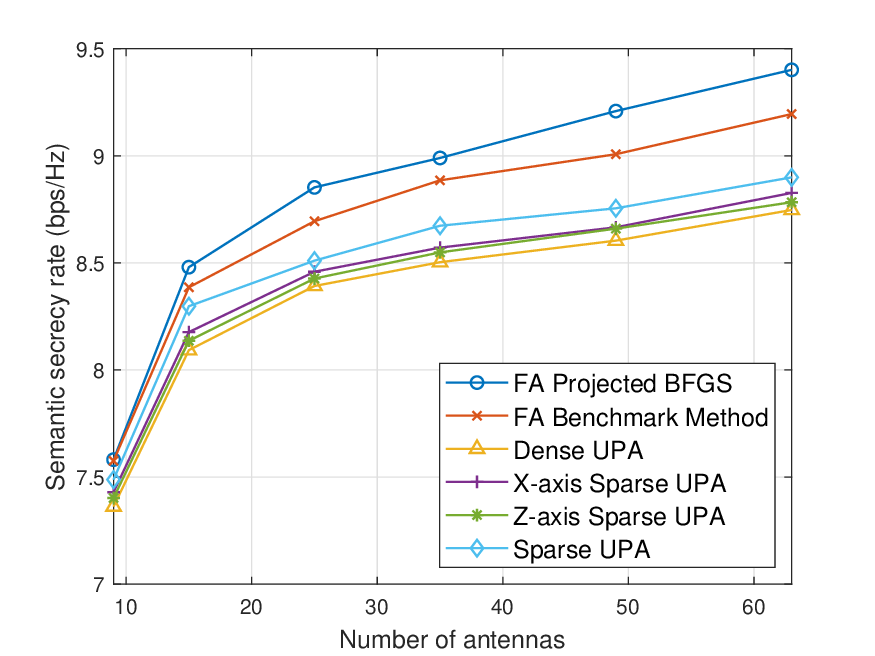}
    \caption{Semantic secrecy rate against number of antennas.}
    \label{fig ssr vs ma}
\end{figure}

Fig.~\ref{fig ssr vs ma} illustrates the semantic secrecy rate performance as a function of the number of antennas for various antenna configurations. Specifically, "Dense UPA" refers to antenna spacing set to half of the wavelength, while "Sparse UPA" denotes antenna spacing equal to $A/8$, with $A$ representing the total antenna array area. It can be observed that the proposed FA Projected BFGS approach consistently outperforms all baseline methods across the entire antenna range. For instance, at approximately 60 antennas, the projected BFGS algorithm achieves a semantic secrecy rate of around 9.4 bps/Hz, compared to about 9.2 bps/Hz attained by the benchmark method. In contrast, conventional array structures, including Dense UPA, X-axis Sparse UPA, Z-axis Sparse UPA, and Sparse UPA, all yield notably lower secrecy rates, although Sparse UPA performs slightly better than other conventional approaches. This improved performance is attributed to Sparse UPA’s reduced channel correlation among users, thereby enhancing secrecy. Moreover, the optimised FA structure allows even more effective exploitation of channel correlation, further boosting performance. Importantly, the performance gap between the proposed FA Projected BFGS method and conventional antenna arrangements widens with increasing numbers of antennas, underscoring the significant advantage of optimised FA structures, particularly in large-scale antenna systems.

\begin{figure}[!t]
    \centering
    \includegraphics[width=0.8\linewidth]{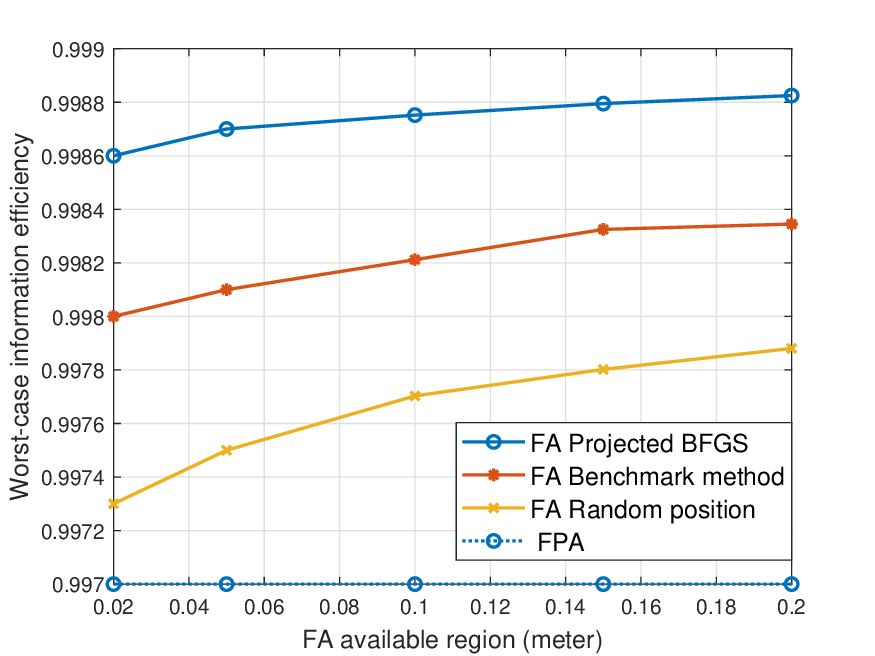}
    \caption{Information efficiency against FAs movable range.}
    \label{fig ie}
\end{figure}

Fig. \ref{fig ie} presents the worst-case information efficiency as a function of the FAs movable region. The result shows that the projected BFGS method achieves the best information efficiency across all values of $\sqrt{C_i}$, which is consistent with the previous results. The benchmark method follows a similar trend but maintains slightly lower information efficiency, highlighting the effectiveness of the projected BFGS method. In contrast, the randomised method shows lower efficiency, while using FPAs results in a flat line with the lowest efficiency throughout, emphasising its limitations due to fixed placement. As $\sqrt{C_i}$ increases, both the projected BFGS and benchmark methods exhibit a saturation behaviour, achieving 0.9989 and 0.9984, respectively. This indicates that while increased antenna mobility enhances performance, the benefit diminishes beyond a certain range. 

\subsection{Sensing Performance}

\begin{figure}[!t]
    \centering
    \includegraphics[width=0.8\linewidth]{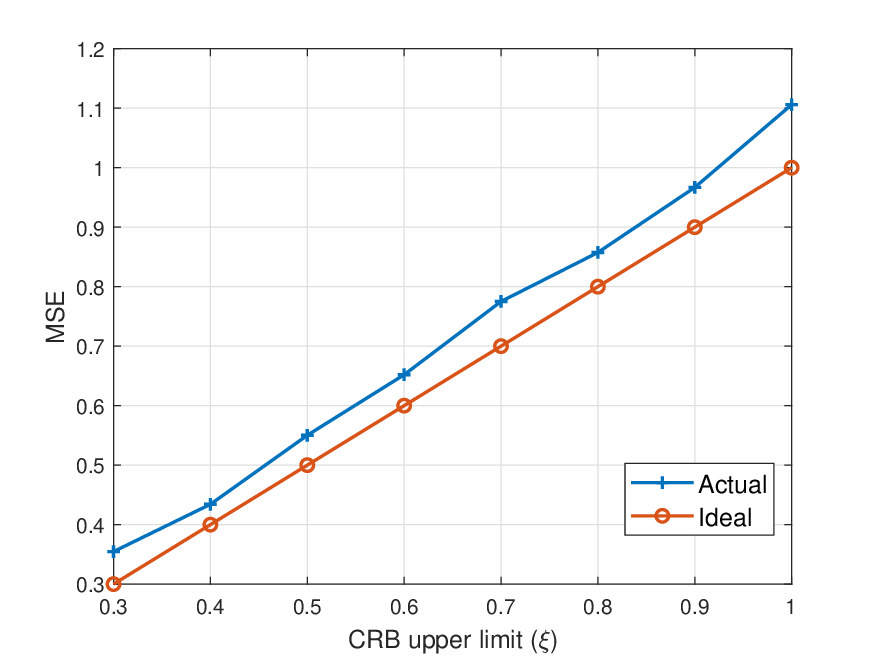}
    \caption{Achieved Mean Square Error against CRB upper bound.}
    \label{CRB MSE}
\end{figure}


Fig.~\ref{CRB MSE} shows the achieved MSE compared to the CRB upper limit ($\xi$). The MSE is calculated using maximum likelihood estimation (MLE). The ideal reference line represents the case where $\text{MSE} = \text{CRB}$, corresponding to the tight theoretical lower bound for a linear model. As $\xi$ increases from 0.3 to 1.0, a noticeable gap appears between the actual MSE and the ideal MSE, indicating that while the estimator closely approaches the theoretical bound, practical limitations, such as the limited number of frames ($F = 100$), introduce a deviation of approximately 10\%, consistent with the standard error ($1/\sqrt{F} = 10\%$). Despite this, the relatively small and stable gap across the entire range of $\xi$ underscores the robustness and effectiveness of the proposed design under varying estimation accuracy requirements.

\begin{figure}[!t]
    \centering
    \includegraphics[width=0.8\linewidth]{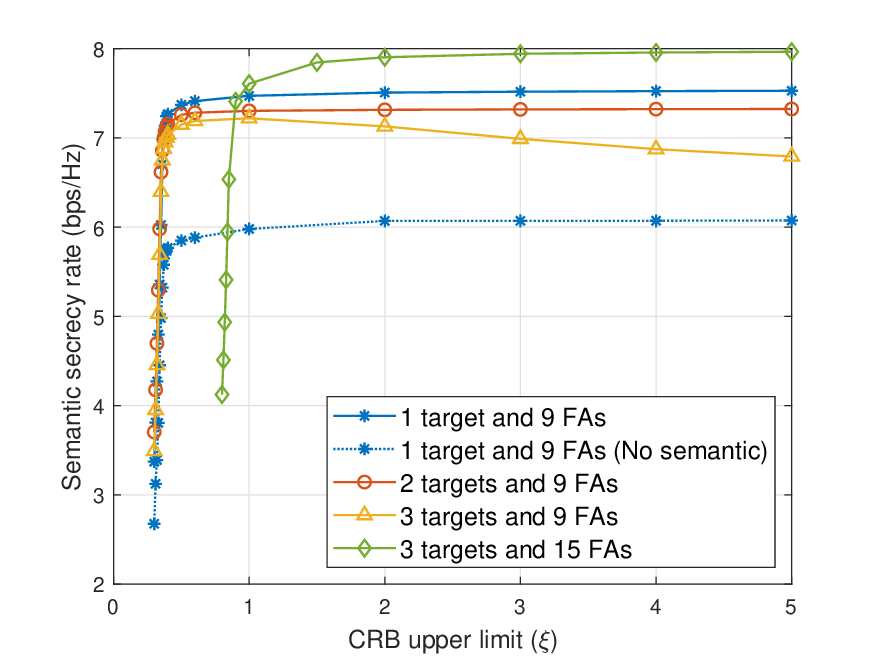}
    \caption{Worst-case semantic secrecy rate against CRB upper bound.}
    \label{crb sr}
\end{figure}

Fig.~\ref{crb sr} illustrates the relationship between the worst-case semantic secrecy rate and the CRB upper limit, denoted as $\xi$, for varying numbers of targets and FAs. The CRB upper limit has been normalised with respect to the antenna size. Notably, under identical $\xi$ values, the design that incorporates semantic (`1 target and 9 FAs') consistently outperforms conventional ISAC designs (`1 target and 9 FAs (No semantic'), highlighting the benefits of semantics in enhancing communication performance without sacrificing sensing accuracy. Furthermore, when the system first achieves an SSR of 6 bps/Hz, the semantic design attains a CRB of approximately 0.4, compared to a CRB of 1 for the non-semantic design. This demonstrates that the semantic design preserves the communication performance of the non-semantic design while significantly enhancing sensing performance.
The figure reveals a rapid increase in the semantic secrecy rate for small values of $\xi$, which subsequently reaches saturation or exhibits a slight decline as $\xi$ increases. This behaviour suggests that, beyond an optimal value of $\xi$, further increases in the CRB upper limit lead to no improvement in terms of communication performance but reduce sensing accuracy. A key observation is the impact of the number of targets. Configurations with fewer targets, such as "1 target and 9 FAs," achieve higher semantic secrecy rates compared to those with more targets, such as "3 targets and 9 FAs." This trend suggests that increasing the number of targets increases the likelihood of communication users being exposed to potential privacy breaches. The number of FAs also plays a critical role. Adding more antennas, as evidenced by the comparison between "3 targets and 9 FAs" and "3 targets and 15 FAs," consistently leads to an enhancement in the semantic secrecy rate. This improvement likely stems from increased spatial diversity in the channels, mitigating the heightened risk of data leakage inherent to configurations with more targets.

\subsection{Computing Performance}

\begin{figure}[!t]
    \centering
    \includegraphics[width=0.8\linewidth]{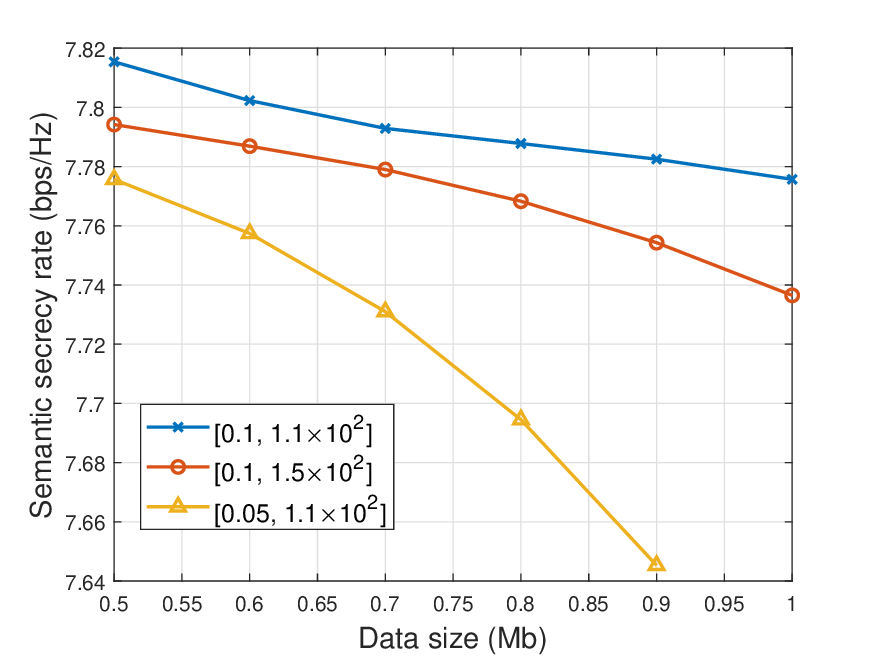}
    \caption{Worst-case semantic secrecy rate against data size per target. The legend $[a,b]$ indicates the maximum tolerance delay of $a$ seconds and the required computational resource $b$ bits/second.}
    \label{ff ssr}
\end{figure}

Fig. \ref{ff ssr} illustrates the relationship between semantic secrecy rate and data size (in Mb) under three different configurations of maximum delay and required computational resources. When the maximum delay is set to 0.1 second and computing capacity of $1.1 \times 10^2$ cycles/bit, the highest semantic secrecy rate is achieved among the scenarios, with a gradual decline as the data size increases. This indicates that a moderate delay and sufficient computational resources are beneficial for maintaining performance. With the same delay but a reduced computing ability of $1.5 \times 10^2$ cycles/bit, there is a slightly reduced communication performance, demonstrating the negative impact of decreased computational ability on the semantic secrecy rate. As the data size increases, the performance difference becomes more obvious. When a stricter delay of 0.05 seconds is used and a computing capacity of $1.1 \times 10^2$ cycles/bit, the lowest semantic secrecy rate is obtained, emphasising the trade-off between achieving low latency and maintaining a high semantic rate within the constraints of limited resources.

\section{Conclusion and Future Direction}

This paper presented a novel joint design for FA-enabled NF-ISCSC systems, considering constraints related to limited computational power and resources. The proposed system enables a BS to simultaneously communicate with multiple CUs and detect multiple extended targets. A joint optimisation problem was formulated to maximise the worst-case semantic secrecy rate while satisfying critical constraints, including sensing accuracy, power consumption, task processing latency, and CPU frequency allocation. To tackle the non-convex optimisation problem, it was decomposed into three sub-problems and solved using an AO approach. The first sub-problem was addressed using a first-order Taylor expansion to ensure convexity and simplify the solution process. For the second sub-problem, a low-complexity projected BFGS algorithm was developed, significantly reducing computational complexity compared to the benchmark method. The third sub-problem was resolved via a search-based approach. Numerical simulations demonstrated the effectiveness of the proposed framework in improving semantic secrecy while maintaining the desired sensing performance within the available computational resources and power budget. These findings highlight the potential of FA-enabled NF-ISCSC systems to support next-generation communication and sensing applications.

While the proposed framework demonstrates promising results, several directions remain open for future research. First, the development of a joint transmit–receive FA structure can be investigated, where transmit-side FAs enhance semantic communication and receive-side FAs facilitate point target sensing tasks. Additionally, examining the impact of antenna positioning on alternative target models, such as the parametric scattering model, presents a valuable research direction. Second, the integration of machine learning techniques could further improve the efficiency and robustness of the proposed framework. Third, extending the system to multi-modal scenarios offers a compelling opportunity to capture richer contextual information and support a broader range of applications. Finally, exploring alternative computing paradigms beyond local processing, such as mobile edge computing (MEC) or over-the-air computation (AirComp), is critical for enhancing the system’s scalability and real-time processing capabilities.

\appendices

\section{Proof and Derivation of Equation \eqref{fim}}
\label{FA TX RX}

\setcounter{equation}{0}
\numberwithin{equation}{section}

We start by considering the linear model described in \eqref{linear model}. In this equation, the matrix \(\mathbf{X}^T \otimes \mathbf{I}_{\left(N_{\text{rx}} \times N_{\text{rz}}\right)}\) is known, whereas the vector \(\mathbf{\bar{g}}\) is deterministic yet unknown. The deterministic property stems from the fact that the antenna positions \(\mathbf{u}\) and \(\mathbf{v}\) are determined and available at the transmitter at the time of echo reception, while the target parameters, though deterministic, remain unknown (as is typical in scenarios such as a Swerling I model, which assumes static or slowly varying targets). Additionally, the noise vector \(\mathbf{\bar{n}}\) follows a Gaussian distribution.

Utilising the established result for linear Gaussian models as detailed in \cite[Chapter 4]{kay1993fundamentals}, the FIM is explicitly given by
\begin{equation}
\begin{aligned}
    \mathbf{J} &= \frac{1}{\sigma^2_r} \left( \mathbf{X}^T \otimes \mathbf{I}_{\left(N_{\text{rx}} \times N_{\text{rz}}\right)} \right)^H \left(  \mathbf{X}^T \otimes \mathbf{I}_{\left(N_{\text{rx}} \times N_{\text{rz}}\right)} \right) \\
    &= \frac{1}{\sigma^2_r} \left(  \left( \mathbf{X}^T \right)^H \mathbf{X}^T \right) \otimes \left( \mathbf{I}_{\left(N_{\text{rx}} \times N_{\text{rz}}\right)}^H \mathbf{I}_{\left(N_{\text{rx}} \times N_{\text{rz}}\right)} \right) \\
    &= \frac{1}{\sigma^2_r} \left(  \mathbf{X}^* \mathbf{X}^T \right) \otimes \mathbf{I}_{\left(N_{\text{rx}} \times N_{\text{rz}}\right)} \\
    &= \frac{1}{\sigma^2_r} \left(  \mathbf{X} \mathbf{X}^H \right)^T \otimes \mathbf{I}_{\left(N_{\text{rx}} \times N_{\text{rz}}\right)} \\
    &= \frac{F}{\sigma^2_r} \mathbf{R}_x^T \otimes \mathbf{I}_{\left(N_{\text{rx}} \times N_{\text{rz}}\right)},
\end{aligned}
\end{equation}
where the final step utilises the covariance relation
\begin{equation}
\mathbf{R}_x = \frac{1}{F} \mathbf{X}\mathbf{X}^H,
\end{equation}
as previously defined in \eqref{covar}.

Thus, the explicit form of the FIM is derived, completing the proof.

\section{Analysis of FIM for a Point Target}
\label{point target}
\setcounter{equation}{0}
\numberwithin{equation}{section}

To analyse the FIM for a point target, we focus on estimating target parameters denoted by \(\zeta \in \{\theta, \phi, d\}\). Consider the radar channel model defined by
\begin{equation}
    \mathbf{G}  = \mathbf{\Psi} \otimes \mathbf{a}^H_r\left(\theta, \phi, d, \mathbf{v} \right) \mathbf{a}_t \left(\theta, \phi, d, \mathbf{u} \right),
\end{equation}
where \(\mathbf{u}\) represents the transmitter FA positions and $\mathbf{v}$ represents the receiver FPA positions. The corresponding vectorised echo signal is given by
\begin{equation}
    \mathbf{\bar{z}}= \Vect\left(\mathbf{Z}\right) = \left( \mathbf{X}^T \otimes \mathbf{I}_{\left(N_{\text{rx}} \times N_{\text{rz}}\right)} \right) \mathbf{\bar{g}} + \mathbf{\bar{n}}.
\end{equation}

Ignoring the path-loss coefficient for simplicity, the FIM for parameters \(\theta, \phi, d\) can be expressed in the structured form
\begin{equation}
    \mathbf{J} = \begin{bmatrix}
        \mathbf{J}_{\theta \theta} & \mathbf{J}_{\phi \theta}  & \mathbf{J}_{d \theta} \\
        \mathbf{J}_{\theta \phi}  & \mathbf{J}_{\phi \phi}  & \mathbf{J}_{d \phi} \\
        \mathbf{J}_{\theta d}  & \mathbf{J}_{\phi d}  & \mathbf{J}_{d d}
    \end{bmatrix},
\end{equation}
with each element defined as
\begin{equation}
    \mathbf{J}_{p, q} = \frac{2 F}{\sigma_r^2} \mathcal{R} \left\{\Tr \left( \frac{\partial \mathbf{\bar{g}}}{ \partial p} \mathbf{R}_x \frac{\partial \mathbf{\bar{g}}^H}{ \partial q} \right) \right\},
\end{equation}
where \(\mathbf{R}_x\) denotes the transmit signal covariance matrix.

To establish the dependence of the FIM \(\mathbf{J}\) on the transmitter FA positions \(\mathbf{u}\), we examine the derivative structure \(\frac{\partial \mathbf{\bar{g}}}{\partial p}\). Recall that \(\mathbf{u}\) comprises FA coordinates \(\{x_i, z_i\}\), and each component \(\bar{g} \in \mathbf{\bar{g}}\) can be explicitly written as
\begin{equation}
\begin{aligned}
    \bar{g} = \exp\Bigg( j\frac{2\pi}{\lambda} \Bigg(&m d_x \cos\theta \sin\phi - \frac{m^2 d_x^2 (1-\cos^2\theta \sin^2\phi)}{2d} \\
    &+ m d_z \cos\phi - \frac{m^2 d_z^2 \sin^2\phi}{2d} \\
    &+ x_i \cos\theta \sin\phi - \frac{x_i^2 (1-\cos^2\theta \sin^2\phi)}{2d} \\
    &+ z_i \cos\phi - \frac{z_i^2 \sin^2\phi}{2d} \Bigg) \Bigg).
\end{aligned}
\end{equation}

For instance, considering the derivative with respect to \(\theta\), each element of \(\frac{\partial \mathbf{\bar{g}}}{\partial \theta}\) is given by
\begin{equation}\label{derivative}
    \frac{\partial \bar{g}}{\partial \theta} = \dot{\bar{g}} \bar{g},
\end{equation}
where \(\dot{\bar{g}}\) denotes the derivative term with respect to \(\theta\).

It is evident from \eqref{derivative} that this derivative explicitly depends on the transmitter FA positions \(x_i, z_i\). Since each element of the FIM incorporates these derivatives, the entire FIM and consequently the CRB, is inherently a function of the FA positions~\(\mathbf{u}\). This argument extends naturally to other FA structures, including Tx-Rx or Rx-only designs.

\section{Convergence Analysis}
\label{coverge Appendix}
\setcounter{equation}{0}
\numberwithin{equation}{section}

Problem~\eqref{opt1} is decomposed into three sub-problems as shown in \eqref{opt2}, \eqref{opt FA}, and \eqref{opt semantic}. Denoting the original optimisation problem as \( M \), and the sub-problems as \( M_1 \), \( M_2 \), and \( M_3 \), respectively, we have:
\begin{equation}
    \max M \triangleq \max M_1 + \max M_2 + \max M_3.
\end{equation}

The sub-problem \( M_1 \) is non-convex with respect to \( \mathbf{w}_k \) and \( \mathbf{R}_x \). We show that \( M_1 \) increases when jointly updating \( [\mathbf{w}_k, \mathbf{R}_x, f_l] \):
\begin{equation}
\begin{aligned}
    M_1^{(e)} &= G_1 \left(\mathbf{w}_k^{(e)}, \mathbf{R}_x^{(e)}, \mathbf{u}_i, f_l, \rho_k \right) \\
    &\overset{(a)}{\leq} G_1 \left(\mathbf{w}_k^{(e+1)}, \mathbf{R}_x^{(e+1)}, \mathbf{u}_i, f_l, \rho_k \right) = M_1^{(e+1)}, 
\end{aligned}
\end{equation}
where \( (a) \) follows from the fact that the surrogate function (first-order Taylor expansion) yields a non-decreasing objective. Since \( M_1 \) is bounded due to power and resource constraints, it is guaranteed to converge.

The sub-problem \( M_2 \) is non-convex with respect to \( \mathbf{u} \). The projected BFGS algorithm, combined with backtracking line search, generates a sequence of iterates \( \{ \mathbf{u}_e \} \) to maximise \( g(\mathbf{u}) \). Since \( g(\mathbf{u}) \) is twice Lipschitz continuously differentiable and locally quadratic in \( \mathbf{u} \), and the approximated Hessian \( (\nabla^2_e)^{-1} \) remains positive definite, the algorithm is guaranteed to converge to a stationary point under standard conditions~\cite{schmidt2009optimizing, nocedal1999numerical}.

The sub-problem \( M_3 \) is a linear program with respect to \( \rho_k \) (when other variables are fixed), and is bounded by box constraints. Hence, it admits a unique optimal solution, and any standard convex optimisation algorithm applied to \( M_3 \) is guaranteed to converge to this solution.

Finally, since each sub-problem is solved to convergence within the alternating optimisation framework, and the overall objective is non-decreasing and bounded, the original problem is guaranteed to converge to a stationary point.

\bibliographystyle{ieeetr}
\bibliography{ref}

\end{document}